\documentclass[aps,prb,twocolumn,floatfix,showpacs]{revtex4-1}
\usepackage{graphicx}
\usepackage{amsmath,bm,amssymb,amsfonts}

\begin{document}

\title{Longitudinal optical and spin-Hall conductivities of Rashba
conducting strips coupled to ferro- and antiferromagnetic layers}

\author{Jos\'e A. Riera}
\affiliation{Instituto de F\'{\i}sica Rosario (CONICET) and
Universidad Nacional de Rosario, Rosario, Argentina
}

\date{\today}

\begin{abstract}
A system composed of a conducting planar strip with Rashba spin-orbit
coupling (RSOC), magnetically coupled to a layer of localized
magnetic moments, at equilibrium, is studied within a microscopic
Hamiltonian with numerical techniques at zero temperature in the clean
limit. In particular, transport properties for the cases of
ferromagnetic (FM) and antiferromagnetic (AFM) coupled layers
are computed in linear response on strips of varying width. Some
behaviors observed for these properties are consistent with the
ones observed for the corresponding Rashba helical currents.
The case of uncoupled Rashba strips is also studied for comparison.
In the case of Rashba strips coupled to an AFM localized order,
results for the longitudinal dc conductivity, for small strip widths, 
suggest the proximity to a metal-insulator transition. More
interesting, in the proximity of this transition, and in general
at intermediate values of the RSOC, it is observed a large spin-Hall
conductivity that is two orders of magnitude larger than the one for
the FM order for the same values of the RSOC and strip widths. There
are clearly two different regimes for small and for large RSOC,
which is also present in the behavior of Rashba helical currents.
Different contributions to the optical and the spin-Hall
conductivities, according to a new classification of inter- or
intra-band origin proposed for planar strips in the clean limit, or
coming from the hopping or spin-orbit terms of the Hamiltonian, are
examined. Finally, the effects of different orientation of the
coupled magnetic moments will be also studied.

\end{abstract}

\maketitle

\section{Introduction}
\label{introsection}

There is currently an increasing interest in studying and developing new
systems and devices that could process information using the spin of the
electron, which is the essence of the field of spintronics.\cite{wolf,
prinz,zutic,awschalom}
In particular, a considerable number of possibilities stem from the
implementation of effective couplings derived from microscopic
spin-orbit (SO) interactions, chief among them the Rashba spin-orbit 
coupling (RSOC) which appears in systems with structural inversion
asymmetry.\cite{rashba,winkler,sinovaRMP,hoffmann13,bercioux}

It has been recently noticed\cite{manchon08,manchon09,miron10,xwang}
that a strong spin torque can be induced on ferromagnets (FM) coupled
to a two-dimensional (2D) layer with Rashba SOC. This process was
observed when an electrical current flows in the plane of a Co layer
with asymmetric Pt and AlOx interfaces.\cite{miron10,pesin,freimuth}
Even more recently, it has been discussed the possibility of an
analogous relativistic SO torque appearing when an antiferromagnetic
(AFM) layer is coupled to a conducting layer containing
RSOC.\cite{nunez,sinova14} It was also suggested that this possibility
could be realized in bulk Mn$_2$Au, which although is centrosymmetric,
it can be divided into two sublattices that separately have broken
inversion symmetry. This second
possibility is referred to as a N\'eel SO torque (NSOT),
to differentiate it from the previously mentioned FM SO torque (FSOT).
In these two systems, the FM or AFM order in the magnetic layer is
fixed due to a large enough exchange interaction among localized
magnetic moments. Possible advantages of spintronic devices
involving AFM layers, for example their fast magnetic dynamics and
the insensitivity to stray fields, are well-known.\cite{roy2016,
wadley16} Remarkably, it was recently found that the NSOT could drive
an antiferromagnetic domain wall at velocities two orders of magnitude
greater than the ones in ferromagnets.\cite{gomonay,zelezny16}

Rashba SO coupling leads to the spin-Hall effect
\cite{sinovaRMP,hoffmann13} which manifests itself on finite width
systems or strips, as spin accumulation at the strip edges.
\cite{kato,jungwirth,riera}
Perhaps the most important quantity related to these effects is the 
spin-Hall conductivity which in clean 2D systems turned out to be
a constant independent of the Rashba strength for a wide range of
electron fillings.\cite{sinova04} Subsequent studies lead to the 
conclusion that the spin-Hall conductivity vanishes in the presence
of scattering.\cite{rashba04} This result led in turn to some 
controversial interpretation of experimental data. Most of this
theoretical work has been performed on unbounded 2D systems and using
a parabolic band, that is, with infinite bandwidth. Many shortcomings
and peculiar results obtained for these systems have been stressed
in the literature. Specifically, the above mentioned controversy
concerning the spin-Hall conductivity was settled down once the
effect of parabolic bands was recognized, that is, the spin-Hall
conductivity is finite and in general proportional to the square
of the Rashba SOC, as soon as the kinetic energy term departs 
from the parabolic form.\cite{krotkov} On the
other hand, planar strips or wires, rather than unbounded 2D systems,
and tight-binding bands with finite bandwidth, are in general more
realistic, particularly for new materials or mechanisms which
have been proposed for spintronic devices and where larger electron
fillings may be involved, such as for example, those involving
LaAlO$_3$/SrTiO$_3$ interfaces.\cite{hwang,banerjee} It should also
be noticed that on planar strips, at equilibrium, using second
quantization, which allows wave-functions to be finite at the edges,
the presence of Rashba helical currents close to the strip edges have
been reported.\cite{hamad}

Hence, the purpose of the present work is to study the optical and 
spin-Hall conductivities of conducting strips with a Rashba SOC in 
contact with a magnetized slab with FM or AFM orders. Although
the above mentioned FSOT and NSOT occur in off-equilibrium
systems, the study of an appropriate microscopic model Hamiltonian
in second quantization at equilibrium, and within linear response,
could shed some light on the behavior of systems in off-equilibrium
regimes, particularly for strips. Certainly, the simplified model
here considered does not capture many details at the interfaces of
the actual devices\cite{freimuth,sinova14,banerjee}, but this
simplicity is necessary to provide general insights that could
help the search for new materials or devices. These microscopic
insights on the various physical properties examined refer in the 
first place to the hopping or Rashba SO origin of the involved
currents, and in second place, to a classification here proposed
for the energy subbands or modes as a function of momentum that are
characteristic of strips or wires. This classification of
energy-momentum points, discussed in
Section~\ref{isostrip}, leads in turn to a classification of 
inter- and intra-band transitions, which should not be confused
with the ones used in theoretical studies for the unbound plane,
in non-equilibrium regimes and in the presence of
disorder (See for example, Ref.~\onlinecite{hli2015}).

For the sake of comparison, the
case of isolated Rashba conducting strips, that is, not connected to
a magnetic layer, will also be studied. It is also interesting 
to examine the Rashba helical currents, and to correlate their
behavior is correlated with the one observed
through the optical and spin-Hall conductivities. These studies are
performed assuming that the magnetic moments are always collinear.
This assumption is realistic since the exchange interaction between
moments in the coupled layer is typically
very larger.\cite{zelezny16} In the final Section, the effects of
various possible orientations of the coupled magnetic order, for both
the FM and AFM layers, will be examined, particularly for the AFM
case when the magnetic moments are oriented along a direction that
is parallel to the conducting plane, which is the situation 
considered in Ref.~\onlinecite{sinova14}.

\section{Model and methods}
\label{modelset}

The Hamiltonian here studied is 
$H = H_{0} + H_{int}$ where \cite{meza,sinova14}:
\begin{eqnarray}
H_{0} &=& - t \sum_{<l,m>,\sigma} (c_{l\sigma}^\dagger c_{m\sigma} +
    H. c.) + \lambda \sum_{l}
     [c_{l+x\downarrow}^\dagger c_{l\uparrow}  \nonumber  \\
  &-& c_{l+x\uparrow}^\dagger c_{l\downarrow} + i (
   c_{l+y\downarrow}^\dagger c_{l\uparrow}
  + c_{l+y\uparrow}^\dagger c_{l\downarrow}) + H. c.] \nonumber  \\
H_{int} &=& - J_{sd} \sum_{l} {\bf S}_l \cdot {\bf \hat{s}}_l
        + J \sum_{<l,m>} {\bf S}_l \cdot {\bf S}_m
\label{ham1orb}
\end{eqnarray}
where $l,m$ are sites on a square lattice, located on the 
$\{x,y\}$-plane, ${\bf S}_l$ are the
localized magnetic moments, assumed classical, and
${\bf \hat{s}}_l$ are the spin of the conduction electrons
(its operator nature is made explicit for later usage).
The longitudinal (transversal) direction of the strip corresponds
to the $x$ axis ($y$ axis). $H_{0}$ is the noninteracting part, which
includes the hopping and RSOC terms with coupling constants $t$ and
$\lambda$, respectively. The RSOC term corresponds to an effective
Rashba electric field along the $z$-axis, i.e. perpendicular to the
plane of the strip. Since both terms in $H_{0}$ contribute to
the total kinetic energy, we choose the normalization
$t^2+\lambda^2=1$, whose square root will then be adopted as the
unit of energy. With this normalization, the kinetic energy, and
hence the total energy for fixed $J,~J_{sd}$, turns out to be
approximately constant as $\lambda/t$ is varied.\cite{meza} Hence,
all the physical properties studied in the following will solely
depend on the ratio $\lambda/t$, for given values of $J, J_{sd}$.
The interacting part of the Hamiltonian contains a ferromagnetic
coupling between conduction electrons and localized magnetic moments
with strength $J_{sd}$, and an exchange interaction between magnetic
moments with coupling $J$. Hamiltonian (\ref{ham1orb}) is just a
ferromagnetic Kondo lattice model with a Rashba SO coupling.

For the FM case, for most of the calculations, a value of $J_{sd}=10$
will be adopted, which
corresponds to the case of well-separated spin-up and spin-down
conduction bands. For $J_{sd}=5$, the bands are partially separated
and in this case an AFM or staggered order along the $y$ direction
would be energetically favourable if the magnetic moments were allowed
to rotate, for $J=0$.\cite{meza2} For this reason
the AFM order will be studied for $J_{sd}=5$, although some results
for $J_{sd}=10$ will be also presented. In general, results do not
qualitative change for this case in this range of $J_{sd}$.
For the FM case, some
results obtained by varying $J_{sd}$ will be discussed in
Section~\ref{coupledfm} though.
Since FM and AFM orders are put by hand, the value of $J$ is
irrelevant since it only adds a constant to the total energy.
The direction of the magnetic moments is adopted to be along the
$z$-axis. Since model (\ref{ham1orb}) corresponds to a quantization
axis along the $z$-axis, it is expected that results will depend on
the direction of the magnetic moments, and this issue will be examined
in the final section.

Hamiltonian (\ref{ham1orb}) will be studied on strips of length $L$
and width $W$, with periodic (open) boundary conditions along the
longitudinal (transversal) direction. $N=LW$, is the number of
sites on the strip.

For classical localized magnetic moments, which is the assumption
usually adopted in this context, the Hamiltonian becomes an extended
tight-binding problem which is solved by numerical exact
diagonalization for clusters 
with $2\le W \le 64$ and $256 \le L \le 8000$. In all results
presented below, finite size effects with respect to $L$ are
negligible. All results presented below correspond to quarter
filling, $n=0.5$.

The main quantities studied are the spin-conserving currents,
$J_{\sigma,\hat{\mu}}$, $\sigma=\uparrow,\downarrow$,
$\hat{\mu}=x, y$, defined as the expectation value of the operator:
\begin{eqnarray}
\hat{j}_{\sigma,l,\hat{\mu}} = i t
(c_{l+\hat{\mu},\sigma}^\dagger c_{l,\sigma} - H. c.),
\label{curhop}
\end{eqnarray}
in units where the electron charge $e=1$, and $\hbar=1$,
and the spin-flipping currents, $J_{SO,\hat{\mu}}$, which are the
expectation value of the operators:
\begin{eqnarray}
\hat{j}_{SO,l,\hat{x}}&=& - i \lambda (c_{l+x\downarrow}^\dagger
   c_{l\uparrow}  -c_{l+x\uparrow}^\dagger c_{l\downarrow} - H. c.)
\nonumber  \\
\hat{j}_{SO,l,\hat{y}}&=& \lambda (c_{l+y\downarrow}^\dagger
   c_{l\uparrow} + c_{l+y\uparrow}^\dagger c_{l\downarrow} + H. c.)
\label{curso}
\end{eqnarray}

These expressions can be obtained in an standard way by introducing
appropriate Peierls factors in the Hamiltonian (\ref{ham1orb}) and
taking the first derivative with respect to the magnetic flux. 
These currents also satisfy the charge conservation law, given
generically by
$\nabla \cdot {\bf \hat{j}}=-\partial n/\partial \tau$, where $n$
is the occupation number operator and $\tau$ is the time, as
long as $\partial n/\partial \tau=i[H_{0},n]$, which will be 
further discussed below.

In equilibrium, due to translation invariance along $x$, all
currents along $x$ depend only on the chain position given by $y$.
In all the cases analyzed below, in the absence of external
electromagnetic sources, the SO currents along the strip direction are
zero, except for the case of Rashba strips connected to a FM layer
as it will be discussed in Section~\ref{coupledfm}. In all cases,
vertical currents are also zero, except for the case of a constant
orientation of the magnetic moments forming an angle $\theta=\pi/3$
considered in Section~\ref{afm_ins}, where a more complex pattern
appears. Certainly, the sum over the strip section of charge currents,
as well as the sum over the strip section of spin currents, is always
zero.

\section{Rashba helical currents}
\label{rhcsection}

The Rashba helical currents (RHCs), are counter-propagating spin-up
and spin-down electron currents, that is of hopping or spin-conserving
nature, at each link at the lattice,\cite{hamad} and they appear due
to the RSOC acting on both $x$ and $y$ directions on the strip,
in equilibrium and in the absence of any external electromagnetic
field. It should be emphasized that they appear as a consequence of
some boundary imposed on the system, or eventually in the
presence of impurities.\cite{bovenzi} Their existence can be inferred
at an effective level\cite{hamad} or by the structure of the
RSOC.\cite{bovenzi,caprara} Of course, the total current along the
strip is zero. In the following we will compute RHCs, defined by the
spin-up current along $x$,
$J_{\uparrow}$, on each chain of closed strips of width $W$.

\begin{figure}[t]
\includegraphics[width=0.8\columnwidth,angle=0]{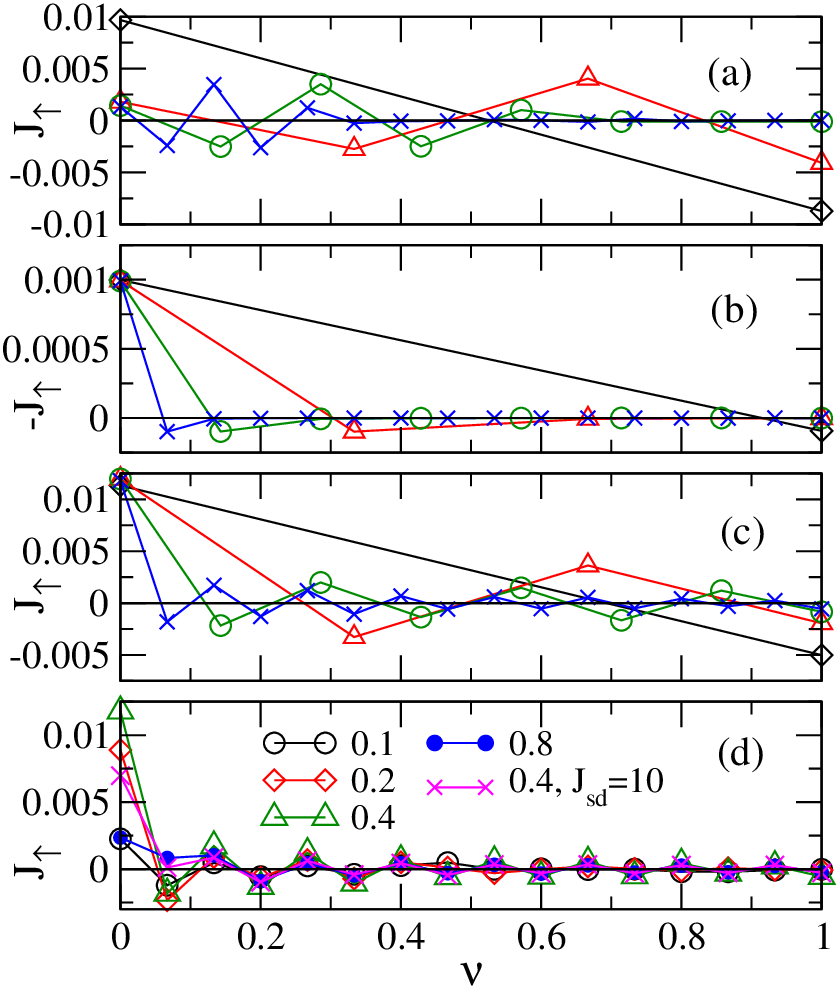}
\caption{(Color online) Spin-up current on each chain as a function of
the depth of the chain ($\nu=0$, edge, $\nu=1$, center chain),
for $\lambda/t=0.4$, (a) isolated Rashba conducting strip
($J_{sd}=0$), (b) FM layer, $J_{sd}=10$, and (c) AFM layer,
$J_{sd}=5$. Results correspond to strips with width $W=4$
(diamonds), 8 (up triangles), 16 (circles), and 32 (crosses).
In (d), Spin-up currents are shown for various values of 
$\lambda/t$ indicated on the plot, AFM layer, $J_{sd}=5$ 
(except the one corresponding to $J_{sd}=10$), $W=32$.
}
\label{fig1}
\end{figure}

The RHCs on each chain as a function of its distance $\nu$ to the edge
($\nu=0$, edge chain ($y=W$), $\nu=1$, center chain ($y=W/2+1$)) are
shown in Fig.~\ref{fig1}(a-c) for
various strip widths, $\lambda/t=0.4$, and different orderings of
the magnetic layer. Fig.~\ref{fig1}(a) corresponds to conducting
strips decoupled from the magnetic layer, that is $J_{sd}=0$.
This is the case previously studied, presenting characteristic
sign oscillations with wavevector $\pi$, at $n=0.5$, and mostly
concentrated near the edges.\cite{hamad} In Fig.~\ref{fig1}(b), the
conducting strip is coupled to a FM layer by a large value of
$J_{sd}=10$, corresponding to well-separated $s^z > 0$ and
$s^z < 0$ bands, where $s^z=<\hat{s}^z>_{sp}$, $<...>_{sp}$ meaning
the average over a single-particle state. This is not a fully
polarized system in the sense that for each band, $|s^z| < 1/2$.
In this case, RHCs are only noticeable from zero at the strip edge,
and effects of finite strip width are negligible. Sign oscillations
are mostly absent. A very different behavior is observed for the case
of an AFM spin background, as shown in Fig.~\ref{fig1}(c). In this
case, RHCs are also maximal near the edge. At the edge,
$J_{\uparrow}(\nu=0)$ is more than an order of magnitude larger than
the one for the FM case, and it does not depend on the strip width.
Besides, similarly to the decoupled system, they present sign
oscillations as a function of $\nu$. Notice also that the direction
of $J_{\uparrow}(\nu=0)$ is the same as the one for the pure Rashba
strip and is opposite to the one of the FM coupled layer case.

The dependence of RHCs with $\lambda/t$ for the AFM background,
for $W=32$ and $J_{sd}=5$, is shown in 
Fig.~\ref{fig1}(d). It is apparent that $J_{\uparrow}(\nu)$ grows 
with $\lambda/t$ up to $\lambda/t\approx 0.4$, in the same
way that it was predicted and observed for the case of isolated
conducting strips,\cite{hamad} and then it starts to decrease for
larger $\lambda/t$. By increasing $J_{sd}$, the RHCs decreases, as
is illustrated for $\lambda/t=0.4$ and $J_{sd}=10$.

\begin{figure}[t]
\includegraphics[width=0.9\columnwidth,angle=0]{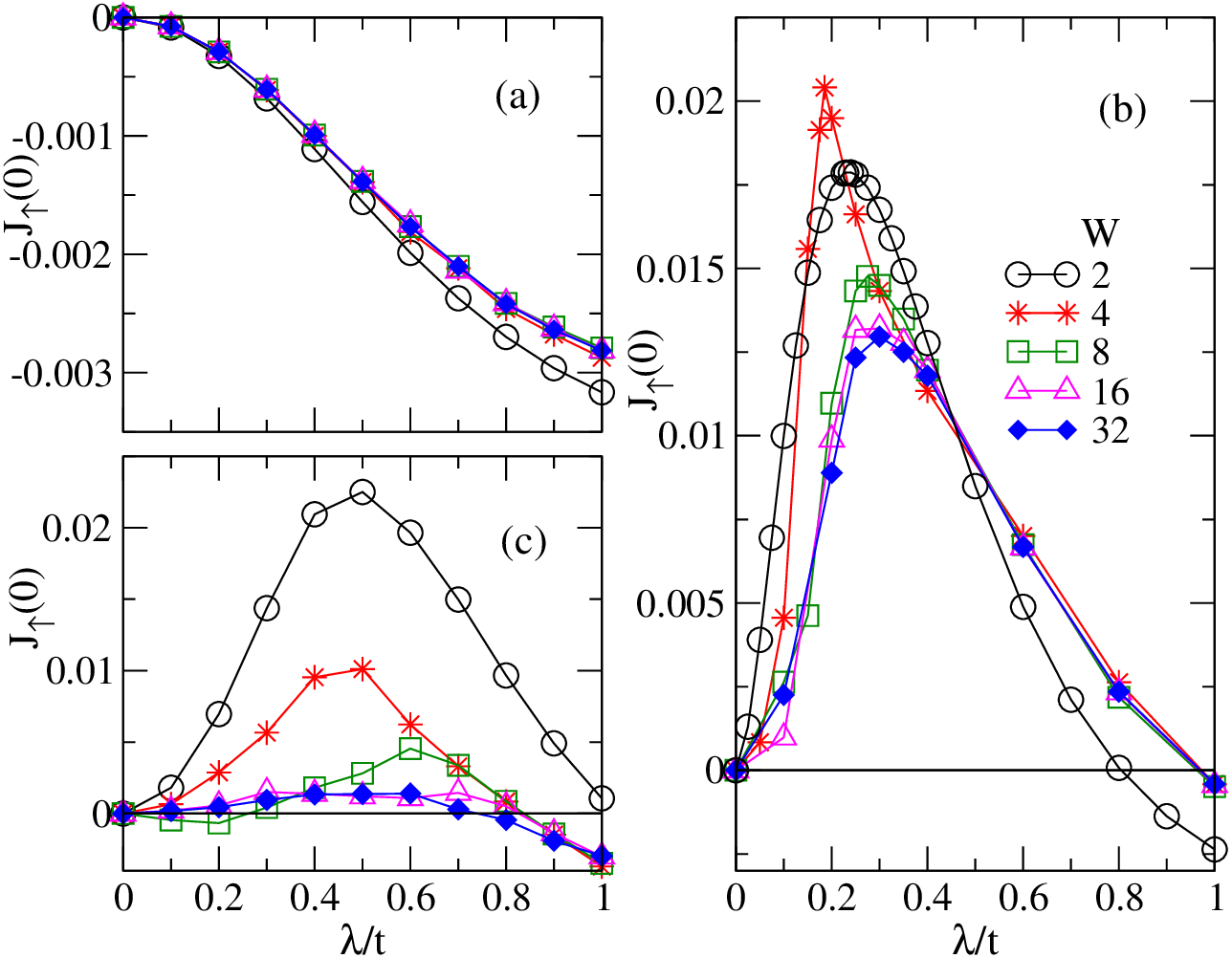}
\caption{(Color online) Spin-up current on the edge chain as a function
of $\lambda/t$, for various strip width $W$ indicated on the
plot, (a) fixed FM state $J_{sd}=10$, (b) fixed AFM state $J_{sd}=5$, and
(c) isolated Rashba strip, $J_{sd}=0$.}
\label{fig2}
\end{figure}

For a more systematic study, from now on only the RHCs on the
outermost chains will be considered. In Fig.~\ref{fig2} the values
of the RHC on the edge chain ($\nu=0$) of strips with various widths
are shown for different magnetic coupled layers as a function of
$\lambda/t$. It can be seen in Fig.~\ref{fig2}(a) that for the FM
layer, the edge RHC in absolute value increases monotonically with
$\lambda/t$ in the range considered, with an initial shape which
approximately follows the predicted quadratic dependence for systems
in the presence of a boundary.\cite{hamad} At $\lambda/t \approx0.5$,
$J_{\uparrow}(0)$ experiences a change of curvature. This behavior is
also virtually independent of $W$ except for the smallest width
considered, $W=2$. In contrast, RHCs on the edge chains follow a
non-monotonic behavior for the AFM layer, as shown in
Fig.~\ref{fig2}(b). In this case, $J_{\uparrow}(0)$ presents a strong
peak located at $0.2 \leq (\lambda/t)_{peak} \leq 0.3$ for all $W$.
The dependence with $W$ is nonmonotonic for
$\lambda/t \leq (\lambda/t)_{peak}$, with the largest peak reached
for $W=4$. This peak
becomes smoother and shifts to larger values of $\lambda/t$ as
$W$ increases, although this variation with $W$ seems to have
converged for $W\approx 32$. It can be observed also that the
the RHC curves in the region of $\lambda/t > (\lambda/t)_{peak}$,
fall onto a single curve, that is, they become independent
of the strip width. Eventually $J_{\uparrow}(0)$ decreases to zero
for $\lambda/t \approx 1$ for all $W$. Remarkably, for $\lambda/t$
closer to the peak position, the RHCs for the AFM coupled layer are
one order of magnitude larger than for the FM layer for all $W$, in
agreement with the results shown in Fig.~\ref{fig1}(b,c).

Results for the isolated conducting Rashba strip, $J_{sd}=0$, are 
shown in Fig.~\ref{fig2}(c) for comparison. $J_{\uparrow}(0)$
presents a broad maximum for $\lambda/t \approx 0.5$, that is
larger than the peak of the AFM coupled layer for $W=2$ but 
decreases rapidly by increasing $W$. Again the dependence with
$W$ seems to be converged for $W\approx 32$. In all cases, the
behavior of $W=2$ seems somewhat different than for wider strips.
This can be attributed to the fact that for $W=2$ both chains are
"outermost", while $W> 2$, the RHCs are spread over the strip
section.

\section{Transport properties}

In this section, the main transport properties appearing
as the response to some applied electromagnetic field, and that 
determine the suitability of Rashba strips coupled to magnetic layers
for spintronic applications, that is, the optical conductivity and the
spin-Hall conductivity, are going to be studied. It is also important
to examine if the behavior of the Rashba helical currents observed in
the preceding section can be correlated with the behavior of these
transport properties.

The optical conductivity is defined as the real part of the linear
response to an electric field and can be written as:\cite{fye}
\begin{eqnarray}
\sigma(\omega) &=& D \delta(\omega) + \sigma^{reg}(\omega)
\label{optcond}  \\
&=& D \delta(\omega)+ \frac{\pi}{N} \sum_{n \neq 0}
    \frac{| \langle \Psi_n | \hat{j}_x | \Psi_0 \rangle |^2}{E_n-E_0}
    \delta(\omega - (E_n-E_0))
\nonumber
\end{eqnarray}
where the paramagnetic current along the $x$-direction is defined in
terms of the currents defined in Eqs.~(\ref{curhop},\ref{curso}) as:
\begin{eqnarray}
\hat{j}_x&=& \hat{j}_{hop,x} + \hat{j}_{SO,x} \nonumber \\
\hat{j}_{hop,x}&=& \hat{j}_{\uparrow,x} + \hat{j}_{\downarrow,x}
 \nonumber \\
\hat{j}_{\sigma,x} &=& \sum_{l} \hat{j}_{\sigma,l,x}, 
     ~~~\sigma=\uparrow,\downarrow \nonumber \\
\hat{j}_{SO,x} &=&  \sum_{l} \hat{j}_{SO,l,x}
\label{current}
\end{eqnarray}
The Drude weight $D$ is calculated from the f-sum rule as:
\begin{eqnarray}
\frac{D}{2\pi} = -\frac{\langle H_{0,x} \rangle}{2N} - I_{reg}
\label{drude}
\end{eqnarray}
where $-\langle H_{0,x} \rangle$ is the total kinetic
energy of electrons along the $x$-direction, and
\begin{eqnarray}
I_{reg} = \frac{1}{N} \sum_{n \neq 0}
\frac{|\langle \Psi_n |\hat{j}_x |\Psi_0 \rangle |^2}{E_n-E_0}
\label{intreg}
\end{eqnarray}
is the integral of the regular part of the optical conductivity.
Notice that from Eq.~(\ref{ham1orb}), 
$\langle H_{0,x} \rangle=\langle H_{0,hop,x}\rangle+
\langle H_{0,SO,x}\rangle$. In addition, taking into account the
two contributions to the total current, given by Eq. (\ref{current}),
$I_{reg}$ has a contribution from hopping currents,
$|\langle \Psi_n |\hat{j}_{hop,x} |\Psi_0 \rangle |^2$, another 
contribution from SO currents,
$|\langle \Psi_n |\hat{j}_{SO,x} |\Psi_0 \rangle |^2$, and the
contribution from the cross terms,
$2 Re\{\langle \Psi_n |\hat{j}_{hop,x} |\Psi_0 \rangle
\langle \Psi_0 |\hat{j}_{SO,x} |\Psi_n \rangle \}$.

The spin-Hall conductivity $\sigma_{sH}$ is defined as the
$\omega=0$ limit of the spin-charge transversal response function
given by the Kubo formula, at zero temperature:\cite{rashba04,sinova04}
\begin{eqnarray}
\sigma^{sc}_{xy}(\omega) &=& -i \frac{1}{\pi N} \sum_{n}
      \sum_{m} 
      \frac{\langle \Psi_n | \hat{j}^s_y | \Psi_m \rangle
     \langle \Psi_m | \hat{j}_x | \Psi_n \rangle}{[(E_n-E_m)^2-\omega^2]}
\label{sHcond}
\end{eqnarray}
where $j^s_y$ is the spin current along the $y$-direction. In the
first sum, the summation is performed only over states with energies 
$E_n$ larger than the Fermi energy $E_F$, and in the second sum
only over states with energies $E_m < E_F$.

The definition of the spin current in Rashba systems has been 
extensively discussed\cite{rashba2003,shi2006,an2012} and perhaps this
issue is still unresolved but a physically reasonable and well-defined
expression from the operatorial point of view is the one that is derived
from the spin conservation equation in the absence of external torques:
$\nabla \cdot {\bf \hat{j}^s} + \partial \hat{S}^z/\partial \tau=0$
($\tau$ is the time).
From this expression, the spin current along the transversal direction
can be computed as:
\begin{eqnarray}
\hat{j}^s_{y,l}=-i[H_{y,l},\hat{s}^z_l],
\label{spcurz}
\end{eqnarray}
where $H_{y,l}$ contains the terms in Eq.~(\ref{ham1orb}) involving
sites $l$ and $l+y$. This expression is a stronger form of the 
spin conservation equation written above. Since on-site terms of
the Hamiltonian could not lead to quantities considered as
"currents", then $[H_{y,l},\hat{s}^z_l]$ should be replaced by
$[H_{0,y,l},\hat{s}^z_l]$,\cite{note} and the following
expressions for $\hat{j}^s_y$ are obtained:
\begin{eqnarray}
\hat{j}^s_y&=&\hat{j}^s_{hop,y} + \hat{j}^s_{SO,y},
\nonumber  \\
\hat{j}^s_{hop,y}&=&\frac{1}{2}(\hat{j}_{\uparrow,y} - 
                    \hat{j}_{\downarrow,y}),
\nonumber  \\
\hat{j}^s_{SO,y}&=& - \frac{\lambda}{2} \sum_{l}
   (c_{l+y\downarrow}^\dagger c_{l\uparrow} -
    c_{l+y\uparrow}^\dagger c_{l\downarrow} + H. c.)
\label{spincury}
\end{eqnarray}
This expression for the spin currents, containing two terms, one from
the hopping and another from the Rashba SO terms of the Hamiltonian
(\ref{ham1orb}), is the second quantized equivalent form of the one
considered in previous works formulated in first quantization and 
using a parabolic kinetic energy.\cite{sinova04}

Clearly, from Eqs. (\ref{current}) and (\ref{spincury}), there will
be in principle four independent contributions to $\sigma_{sH}$ which
are computed and studied separately.

\begin{figure}[t]
\includegraphics[width=0.9\columnwidth,angle=0]{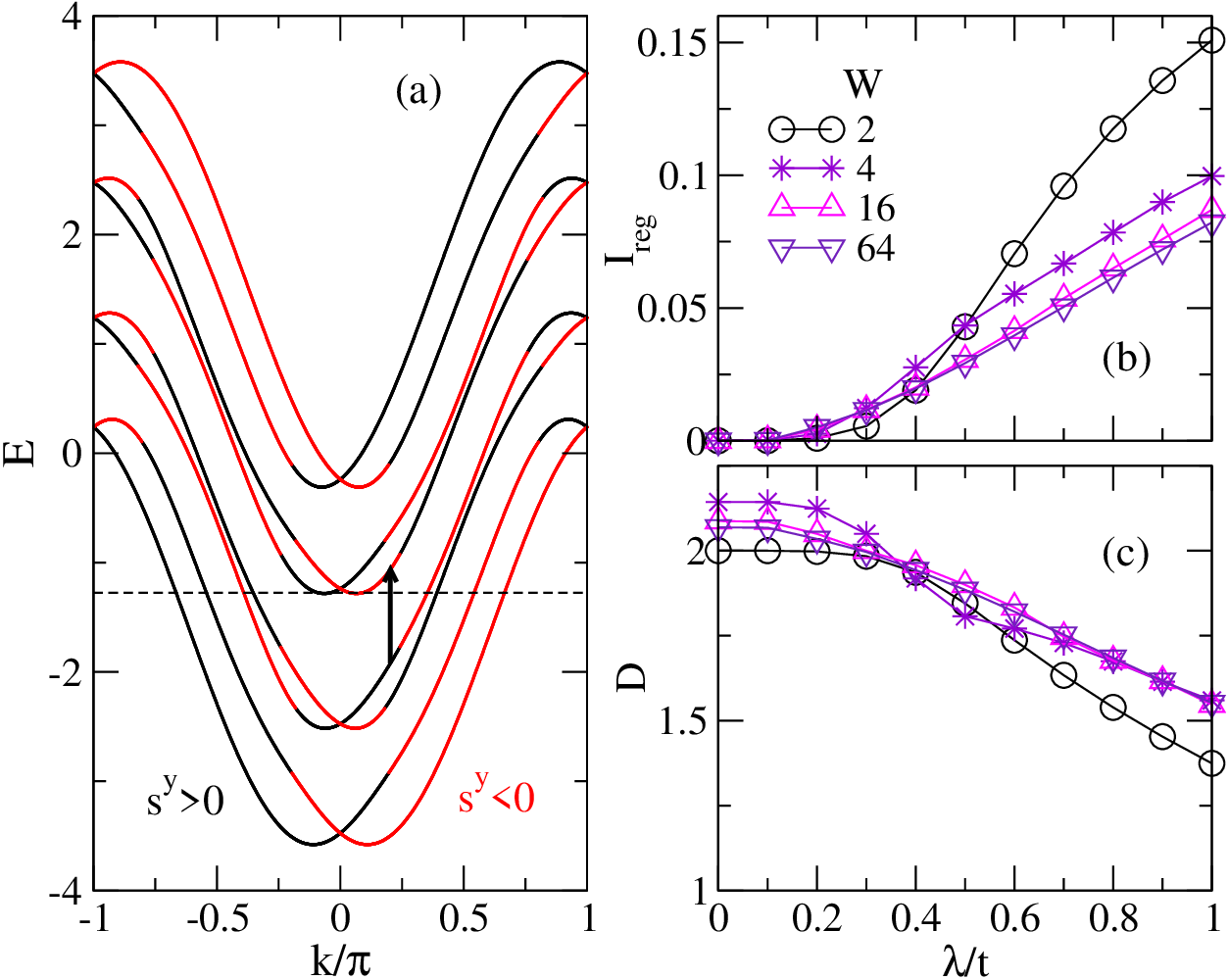}
\caption{(Color online) (a) Energy bands with $s^y>0$ (black) or
$s^y<0$ (red), $W=4$, $\lambda/t=0.4$. $k$ is the momentum along the
longitudinal direction. The dashed line indicates the chemical
potential. (b) Integral of the regular part of the optical
conductivity, and (c) Drude weight,
as a function of $\lambda/t$ for various strip widths $W$.
Isolated, $J_{sd}=0$, $L=2000$ Rashba strips.}
\label{fig3}
\end{figure}

\subsection{Isolated Rashba strips}
\label{isostrip}

Let us start by examining isolated conducting Rashba strips
($J_{sd}=0$). In Fig.~\ref{fig3}(a), the single-particle bands,
$E(k)$, where $k$ is the momentum along $x$, is shown for $W=4$,
$\lambda/t=0.4$. The number of bands is equal to the number of
coupled chains or "modes". The RSOC splits each band into two
subbands with different sign of
$s^y=<\hat{s}^y>_{sp}$.\cite{bercioux,ganichev}
Note that for the strips here considered, as a difference with
the case of the unbounded 2D
system,\cite{ganichev} $s^y$ is not a good quantum number. Also,
due to the breaking of spatial rotational invariance, only
the value of $s^y$, related to the momentum along the strip
axis, is needed to label $\{k,E(k)\}$ points. Note also that
on each of the subbands, the sign of $s^y$ changes.

In the following the expressions "interband transitions" or 
"intraband transitions" will refer to transitions between states
with opposite or the same sign of $s^y$ respectively. This is a
natural extension of the concept of intraband and interband
processes used for the infinite plane\cite{rashba04,sinova04}
to finite-width strips, in the clean limit,
and it also avoids the problem of the presence of the multiple
subbands appearing with strips. More importantly, with this
definition, interband transitions would be detectable by optical
experiments.\cite{ganichev} In the literature, in systems in
a nonequilibrium regime, and in the presence of impurities, other
definitions of inter- and intraband transitions have been 
used.\cite{hli2015}

A typical interband transition contributing to the lowest peak in
$\sigma(\omega)$ is shown in Fig.~\ref{fig3}(a). The inter- and
intraband contributions to $I_{reg}$ and $\sigma_{sH}$ will be
discussed below.

Clearly, the behavior of $E(k)$ is metallic for all electron fillings,
and for $n=0.5$, the chemical potential is located close to a
maximum of the density of states or van-Hove singularity.
In Fig.~\ref{fig3}(b), results for the integral of the regular part
of the optical conductivity, $I_{reg}$, are shown as a function of
$\lambda/t$ for various strip widths indicated on the plot.
The main result is that $I_{reg}$ is due entirely to the SO currents
defined in Eq.~ (\ref{current}).
The Drude weight $D$ is shown in Fig.~\ref{fig3}(c)). Both
quantities present a non-monotonic behavior as a function of $W$,
and cusps around $\lambda/t \approx 0.4-0.5$ can be noticed for
$W=2$ and 4, which is consistent with the behavior of the RHCs shown
in Fig.~\ref{fig2}(c),

\begin{figure}[t]
\includegraphics[width=0.9\columnwidth,angle=0]{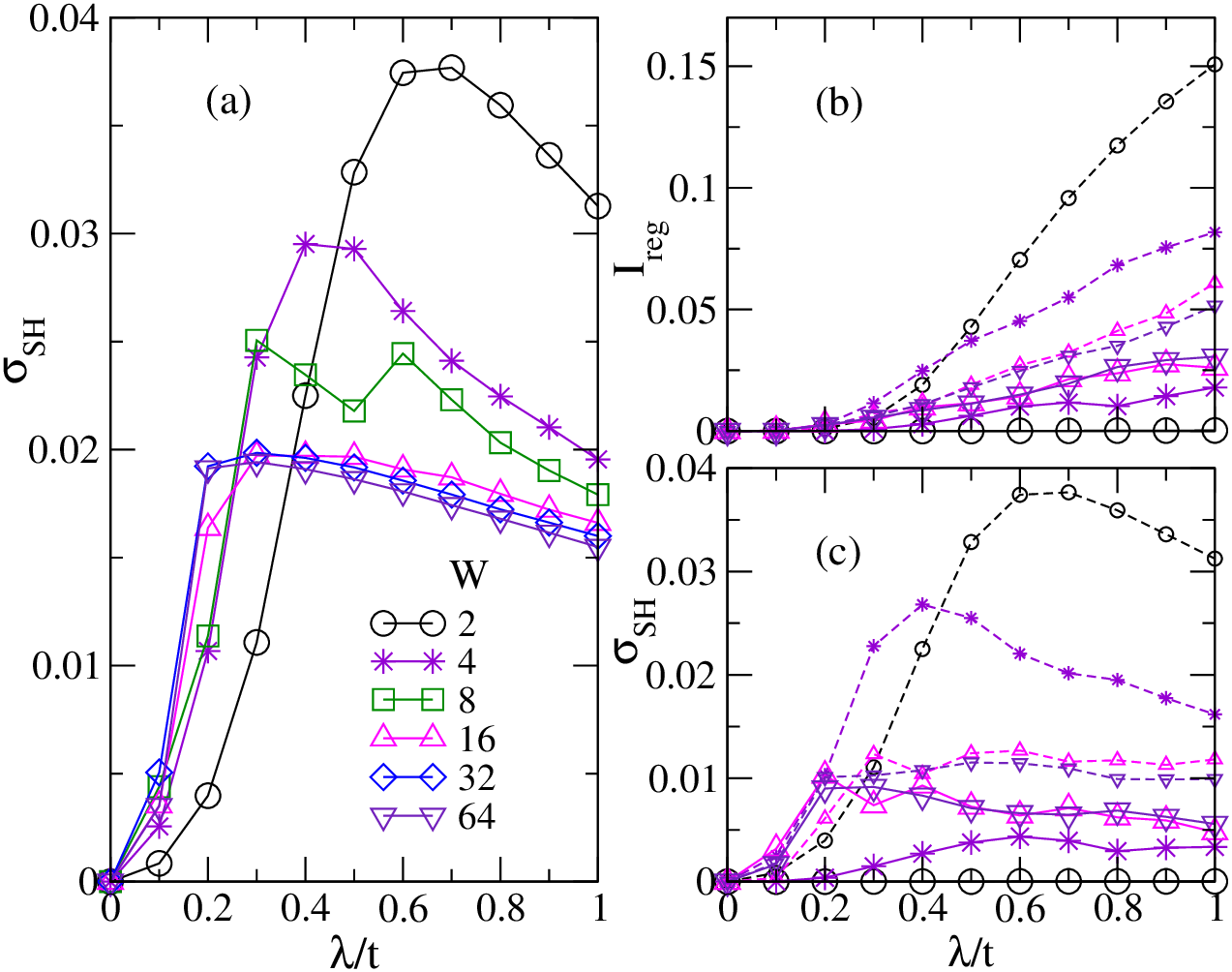}
\caption{(Color online) (a) Total spin-Hall conductivity, (b) inter-
(dashed lines) and intra- (full lines) band contributions to 
$I_{reg}$, and (c) inter- and intra-band
contributions to the spin-Hall conductivity, as a function
of $\lambda/t$ for various strip widths $W$ indicated on the plot.
Isolated, $J_{sd}=0$, $L=2000$ Rashba strips.}
\label{fig4}
\end{figure}

In Fig.~\ref{fig4}(a), the spin-Hall conductivity is shown for
various strip widths as a function of $\lambda/t$. In this case, 
$\sigma_{sH}$ turns out to be entirely due to the
$<\hat{j}_{SO,x}~\hat{j}^s_{hop,y}>$ (for short) contribution.
$\sigma_{SH}$ has a strong dependence with $W$, with a maximum
shifting to lower vales of $\lambda/t$ as $W$ increases,
although it seems that it is converging for $W=64$. For the
largest widths considered, $W=32, 64$, the spin-Hall conductivity
presents a
flat region reminiscent of the constant value predicted in 
Ref.~\onlinecite{sinova04}, although this prediction was obtained
for the infinite 2D system with a parabolic dispersion.
Notice also that with the normalization here adopted for the 
parameters $t$ and $\lambda$, presumably additional factors should
have to be taken into account to compare the present resuls 
with those of Ref.~\onlinecite{sinova04}. For
the widest strips, the presence of a cusp separating two 
different regimes is clearly visible, and in this case, the cusp
is not correlated with the peaks observed in the RHCs curves.

In Figs.~\ref{fig4}(b,c) the inter- and intraband contributions
to the $I_{reg}$ and to $\sigma_{sH}$ are
shown. In both cases, it is remarkable that for $W=2$ both
quantities are purely of inter-band origin, and that as $W$
increases the inter- and intraband processes tend to equally
contribute to those quantities.

\subsection{Coupled FM layer}
\label{coupledfm}

\begin{figure}[t]
\includegraphics[width=0.9\columnwidth,angle=0]{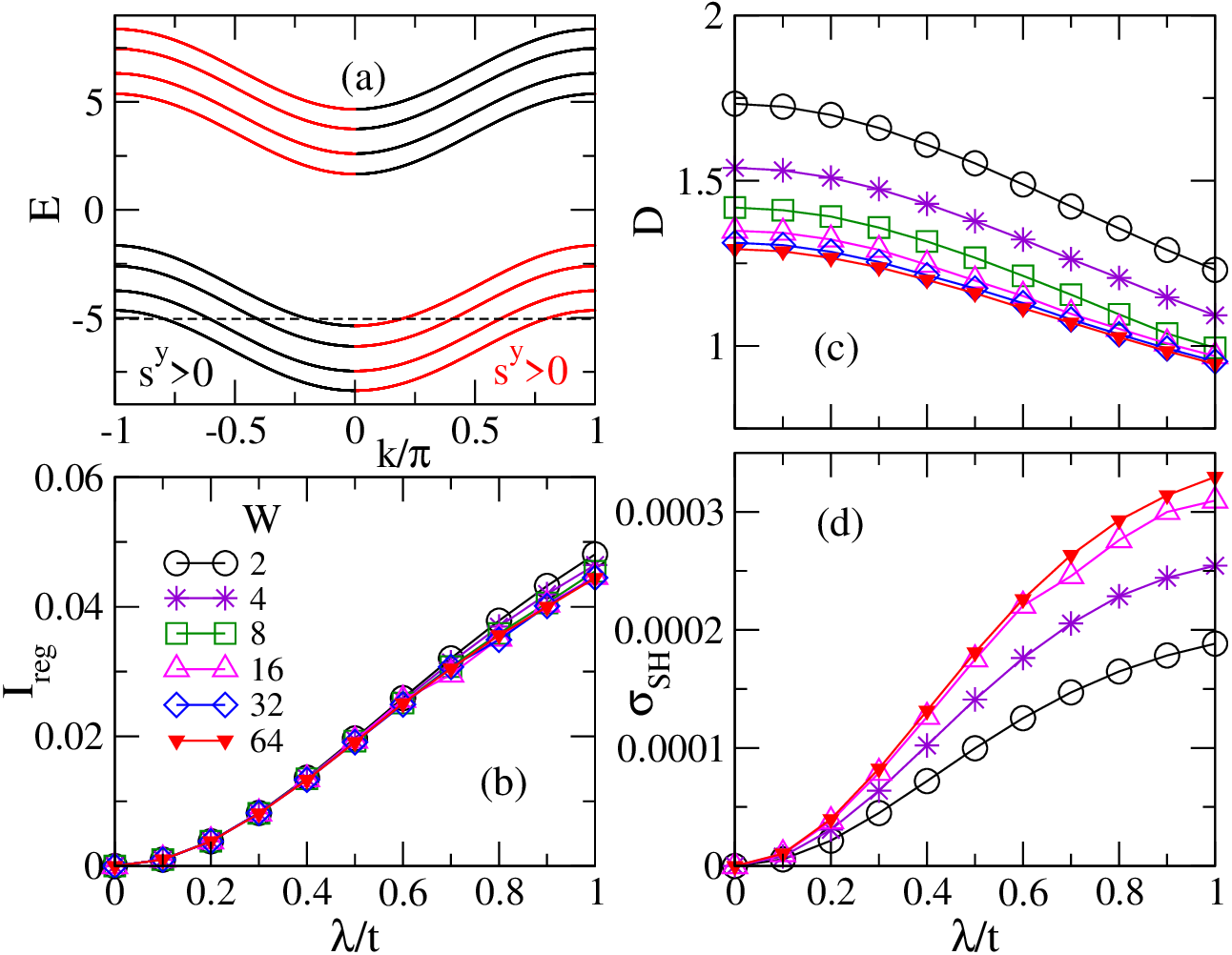}
\caption{(Color online) (a) Band structure for $W=4$, 
$\lambda/t=0.4$. (b) $I_{reg}$, (c) Drude weight, and (d)
spin-Hall conductivity, as a function
of $\lambda/t$ for various strip widths $W$, for a fixed FM
layer, $J_{sd}=10$.}
\label{fig5}
\end{figure}

Let us now consider the more interesting case of a fixed FM layer
coupled to the conducting strip by a ferromagnetic exchange $J_{sd}=10$.
In Fig.~\ref{fig5}(a), the dispersion relation for  $W=4$ and
$\lambda/t=0.4$ is depicted. $E(k)$ points have been labelled by the
sign of $s^y$. The negative energy bands have also $s^z>0$ and are
separated from the positive energy bands with $s^z<0$ due to the
finite value $J_{sd}=10$. The optical conductivity $\sigma(\omega)$
presents a single peak at $\omega\approx 10$ and it is originated
in interband transitions, as can be inferred from the $s^y$ character
of the bands. This will be further discussed below.

In Fig.~\ref{fig5}(b), $I_{reg}$ is
shown as a function of $\lambda/t$ for various strip widths
$W$. Again, this quantity is due entirely to the SO currents.
and it is almost independent of $W$. The Drude weight, shown in
Fig.~\ref{fig5}(c), decreases as $W$ increases due to a reduction
of the kinetic energy along the longitudinal direction.
The vanishing of intraband contributions
to the longitudinal conductivity is consistent with recent results
obtained for an infinite isotropic two-dimensional 
system.\cite{qaiumzadeh}

The spin-Hall conductivity, shown in Fig.~\ref{fig5}(d), also
presents a rather monotonic behavior as a function of $\lambda/t$.
This monotonic behavior in both $D$ and $\sigma_{sH}$ is also
similar to the one observed for the RHCs in Fig.~\ref{fig2}(a),
although there is no direct causality between these features.
The spin-Hall conductivity is also entirely due to the
$<\hat{j}_{SO,x}~\hat{j}^s_{hop,y}>$ contribution, and it
almost entirely involves interband processes, except for marginal
intraband contributions. Notice that in the present case, as well
as in the previous case of isolated Rashba strips, $\sigma_{sH}$
follows an approximate quadratic dependence with the Rashba SOC,
particularly for small $\lambda/t$, consistently with the
prediction of Ref.~\onlinecite{krotkov}.

\begin{figure}[t]
\includegraphics[width=0.9\columnwidth,angle=0]{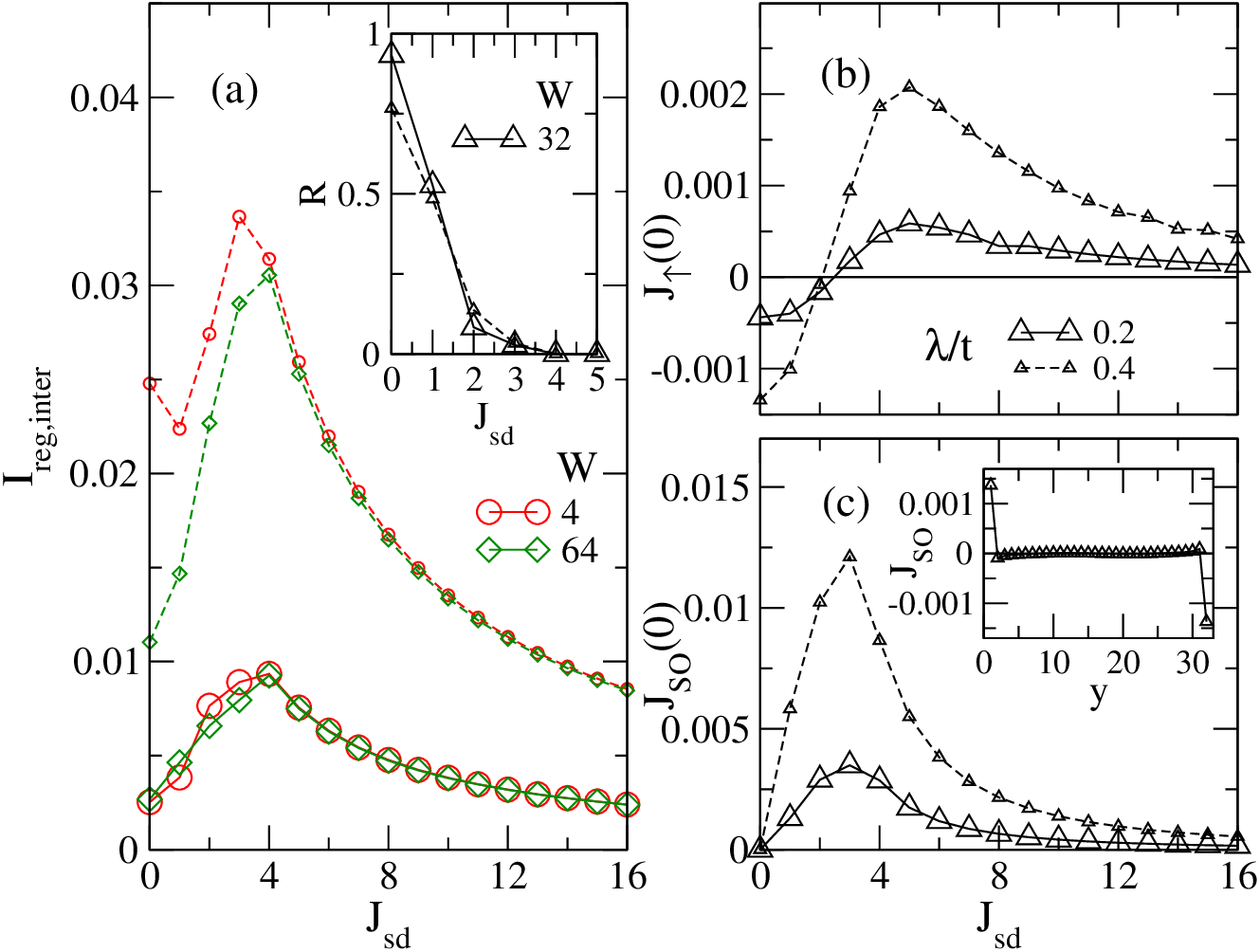}
\caption{(Color online) (a) Interband contribution to $I_{reg}$ 
for $W=4$ (circles) and $W=64$ (diamonds), $\lambda/t=0.2$ and
0.4, as a function of $J_{sd}$. The inset in (a) shows the ratio
$R$ between the intraband and the interband contributions to
$I_{reg}$, $W=32$. (b) Rashba helical currents, and (c) SO
currents on the edge chain, as a function of $J_{sd}$, $W=32$.
The inset in (c) shows $J_{SO}$ on each chain of the $W=32$ strip,
$\lambda/t=0.4$. Coupled FM layer. In all plots, results for
$\lambda/t=0.2$ are shown with full lines, while those for
$\lambda/t=0.4$, with dashed lines.}
\label{fig5bis}
\end{figure}

Taking into account the different behaviors observed for the
isolated Rashba strip, $J_{sd}=0$, and the strip coupled to a FM
layer with $J_{sd}=10$, it is important and interesting to see the
evolution of the various properties examined so far as $J_{sd}$
is varied to zero to a large value. In principle, the behavior for
the FM layer is not extrapolatable to $J_{sd}=0$ due to the
condition of well-separated bands, as shown in Fig.~\ref{fig5}(a).
The results of this study are shown in Fig.~\ref{fig5bis}, and, as
expected, it becomes apparent that this evolution with $J_{sd}$ is
nonmonotonous. It is clear that there is a regime of large
$J_{sd}$, where the $s^z$ up and down bands are well separated,
and a regime 
of small $J_{sd}$, where these bands are partially overlapped.
The crossover between both regimes takes place at 
$J_{sd} \approx 4$. In the region of large $J_{sd}$, $I_{reg}$,
as mentioned above is purely of interband origin and, as it can be
observed in Fig.~\ref{fig5bis}(a), this contribution decays 
approximately as $J_{sd}^{-1}$, as $J_{sd}$ increases, for all
values of $W$ and $\lambda/t$ considered. On the other
hand, $I_{reg,inter}$ decreases as $J_{sd}$ is decreased from
$J_{sd}\approx 4$, and at the same time, as it can be observed in
the inset, the intra-band contribution starts to grow until it 
becomes of the same order as the interband one at $J_{sd}=0$.

The same behavior change can be observed in other quantities. For
example in Fig.~\ref{fig5bis}(b), it is shown that the RHC 
smoothly decreases in the large $J_{sd}$ region as $J_{sd}$ is
increased, and it also decreases as $J_{sd}$ is decreased when
$J_{sd}$ varies from $J_{sd} \approx 4$ towards zero, and it
even change sign in this interval. It is also instructive to
notice that for Rashba strips coupled to a FM layer, not only
Rashba helical currents, which are currents of the hopping type,
are present as in all systems considered in the present work, but
also currents of the spin-flipping type,  $J_{SO}(y)$, along $x$,
defined in Eq. (\ref{curso}), have a finite average on each chain.
One could speculate that these currents appear for the FM coupled
layer because the Rashba SO naturally tends to reduce the FM order,
so the external fixing of a FM order leads to these currents to
counteract that Rashba effect. In fact, if the orientation of the
magnetic moments or the sign of $J_{sd}$ are changed, then these 
helical currents of the SO type change their direction.
The distribution of
these SO currents over the strip section, shown in the inset of
Fig.~\ref{fig5bis}(c) is such that its net value is zero, as it
should be in equilibrium.
Again, as it can be seen in Fig.~\ref{fig5bis}(c), two
different behaviors of $J_{SO}(0)$ can be noticed for $J_{sd}$
greater or smaller than $\approx 4$. Of course, consistently with
the previous results, these spin-flipping currents have vanishing
expectation value for $J_{sd}=0$.

\begin{figure}[t]
\includegraphics[width=0.9\columnwidth,angle=0]{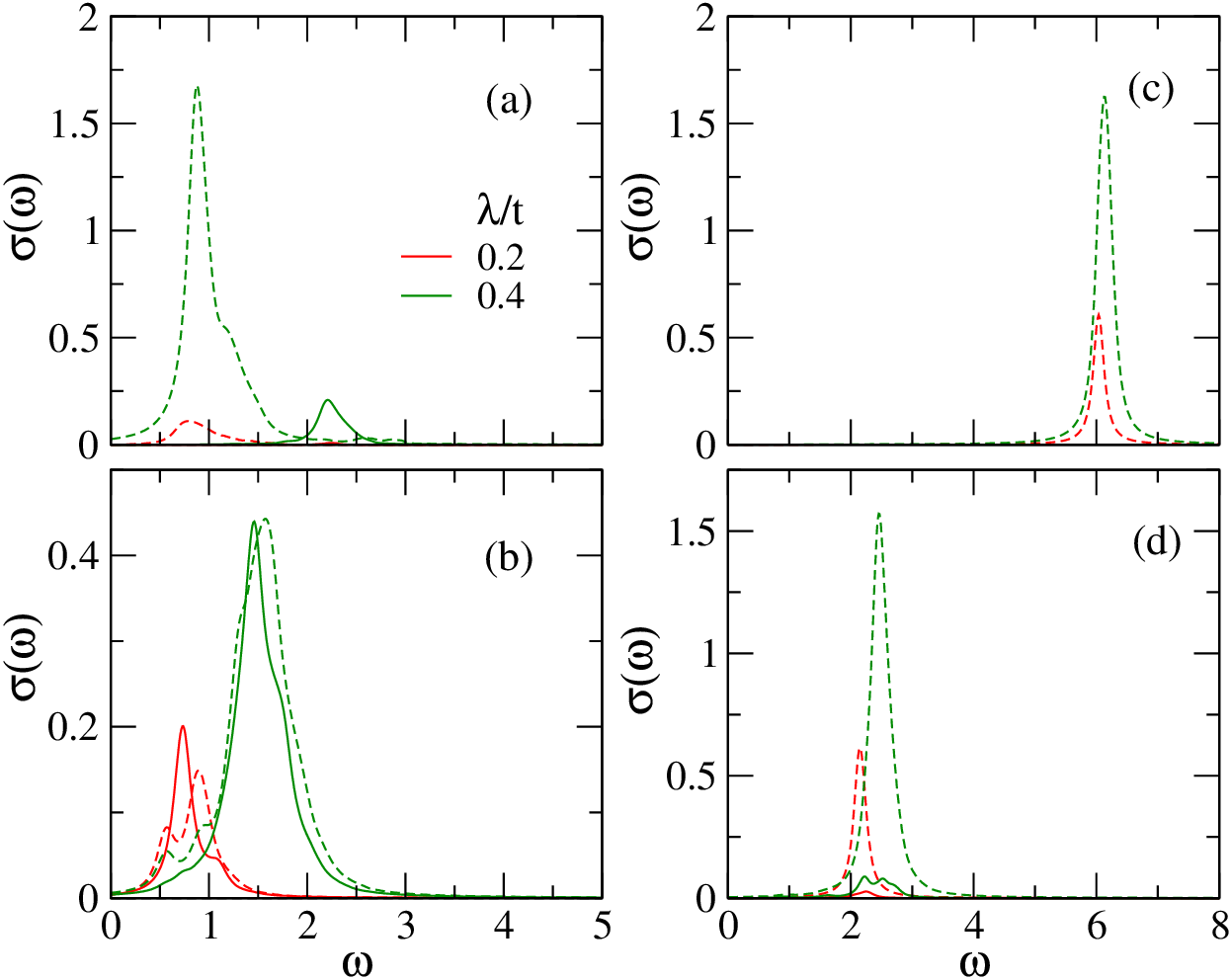}
\caption{(Color online) Optical conductivity as a function of the
frequency for isolated Rashba strips with (a) $W=4$, (b) $W=32$,
and for Rashba strips coupled to a FM layer, $W=16$, for
(c) $J_{sd}=6$, (d) $J_{sd}=2$. Values of $\lambda/t$ are indicated
on the plot. In all plots, the interband (intraband) contribution is
shown with dashed (full) lines, and $\sigma(\omega)$ has been
multiplied by $10^{3}$.}
\label{fig5opt}
\end{figure}

To end this subsection, it is instructive to examine the frequency
dependence of the optical conductivity for some typical cases.
In Fig.~\ref{fig5opt}(a), $\sigma(\omega)$ is shown for the 
isolated Rashba strip for $W=4$, $\lambda/t=0.2$ and 0.4. In this
and related plots, a Lorentzian broadening $\epsilon=0.01$ was
adopted. For the $W=4$ strip, consistently with the results shown
in Fig.~\ref{fig4}, for $\lambda/t=0.2$, $\sigma(\omega)$ is 
almost entirely due to interband transitions, while for
$\lambda/t=0.4$, there is a small contribution from intraband
transitions occurring at higher frequencies. On the other hand,
for a wider strip, $W=32$, Fig.~\ref{fig5opt}(b),the inter- and
intra-band contributions
are approximately equal, and they are originated in transitions
located at roughly the same frequencies. As noticed above regarding
to Fig.~\ref{fig4}, the optical conductivity for isolated strips,
as well as for Rashba strips coupled to FM layers, are entirely
due to SO currents, hence its intensity is always larger for
larger $\lambda/t$. For the case of Rashba strips coupled to FM
layers, in the regime of well-separated bands, or large $J_{sd}$
region, the optical conductivity is purely of interband nature
and it presents a single peak located at $\omega\approx |J_{sd}|$
for any value of $\lambda/t$. This case is shown in
Fig.~\ref{fig5opt}(c) for $J_{sd}=6$, and $W=16$, although
in this region results are mostly independent of $W$. In the
low $J_{sd}$ region, when the bands start to overlap, as it can be
seen in Fig.~\ref{fig5opt}(d) for $J_{sd}=2$, $W=16$, the position
of the largest peak starts to be shifted to to higher frequencies
as $\lambda/t$ increases, and, more importantly, an intraband 
contribution starts to grow. As discussed above, in this small
$J_{sd}$ region, one could expect a monotonous evolution from
the behavior shown in Fig.~\ref{fig5opt}(d) to the one in
Fig.~\ref{fig5opt}(b) as $J_{sd}$ is reduced to zero.

\begin{figure}[t]
\includegraphics[width=0.9\columnwidth,angle=0]{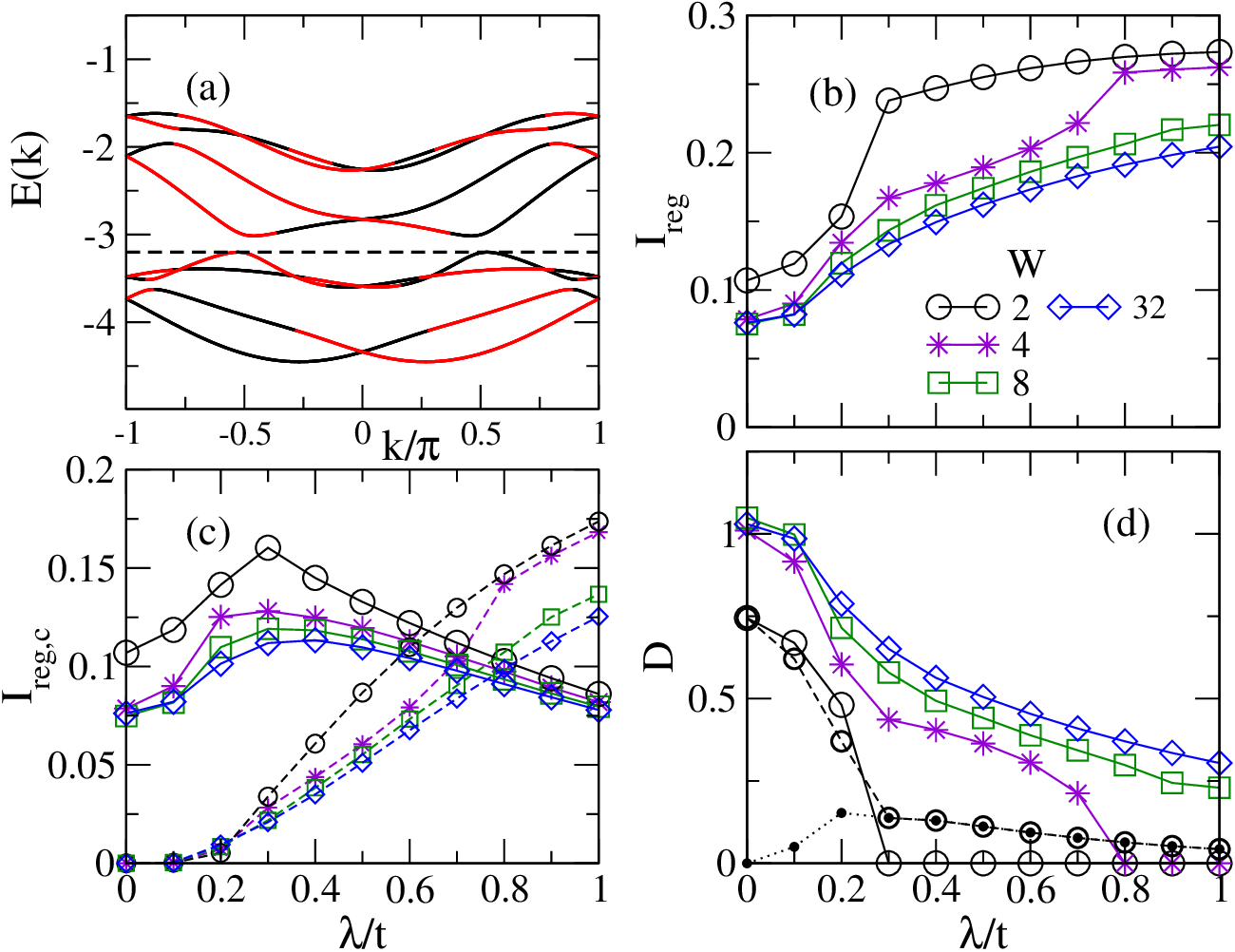}
\caption{(Color online) (a) Lower half of the band structure,
$s^y>0$ (black) or $s^y<0$ (red), $W=4$, $\lambda/t= 0.4$.
(b) $I_{reg}$, and (c) contributions to $I_{reg}$ from hopping
(solid lines) and SO (dashed lines) currents, 
(d) Drude weight, as a function of $\lambda/t$,
for coupled AFM layer, $J_{sd}=5$. Results for $256\times W$ strips
with symbols for various $W$ indicated on the plot.
In (d), contributions
to $D$ from hopping (dashed line) and SO (dotted line) currents
for $W=2$ are also included.}
\label{fig6}
\end{figure}

\subsection{Coupled AFM layer}

Let us finally examine the most important case which is the one of a
Rashba conducting strip coupled to an AFM layer. Fig.~\ref{fig6}(a)
shows the lowest half energy bands of the $W=4$ strip, 
$\lambda/t= 0.4$, where again $E(k)$ points have been
labelled according to the sign of $s^y$. There is another set of
energy bands symmetrically located with respect to $E=0$. The two
set of bands are separated in energy due to the finite $J_{sd}=5$
value. A remarkable difference with respect to the two previous
cases, is that the coupling to a AFM layer introduces small gaps
whose sizes decreases with $W$ and increases with $J_{sd}$. For
$J_{sd}=5$ the gap is present for $W\le8$, and disappears for
$W \ge 16$. For $W\le8$ and $n=0.5$, the Fermi level is just at
the top of the energy branch below the gap, and hence close to
a high density of states.

Fig.~\ref{fig6}(b) shows $I_{reg}$ for various strip widths $W$,
and Fig.~\ref{fig6}(d) the corresponding results for the Drude
weight. For the smallest strip widths, $W=2,4$ there is a cusp 
at $\lambda/t\approx 0.3$, and $\lambda/t\approx 0.8$,
that coincides with the value at which the respective 
Drude weights vanish. Notice that for widths $W >2$, there is also
a curvature change near $\lambda/t=0.2$. It seems then that the
decaying of the RHCs after their peak, shown in Fig.~\ref{fig2}(b),
may be correlated with the onset, for $W=2$, or the proximity to a
metal-insulator transition for $W >2$. It is also interesting to
note that the vanishing of the Drude weight for $W=2,~4$ is due to
equal and opposite contributions from spin-conserving and spin-
flipping currents, as shown in Fig.~\ref{fig6}(d) for $W=2$.

As expected from the complex pattern of
$E(k)$ points with positive and negative values of $s^y$, and as
expected from the already discussed case of free Rashba strips
($J_{sd}=0$), $I_{reg}$ will have contributions from both inter- and
intraband processes. More interesting, and perhaps more physically
appealing, $I_{reg}$ has now contributions from both the SO
(spin-flipping) and the hopping (spin-conserving) currents, as shown
in Fig.~\ref{fig6}(b). As expected, the contribution from
$\hat{j}_{hop,x}$ ($\hat{j}_{SO,x}$) dominates at small (large)
$\lambda/t$, Contributions from the mixed term are marginal. 

\begin{figure}[t]
\includegraphics[width=0.9\columnwidth,angle=0]{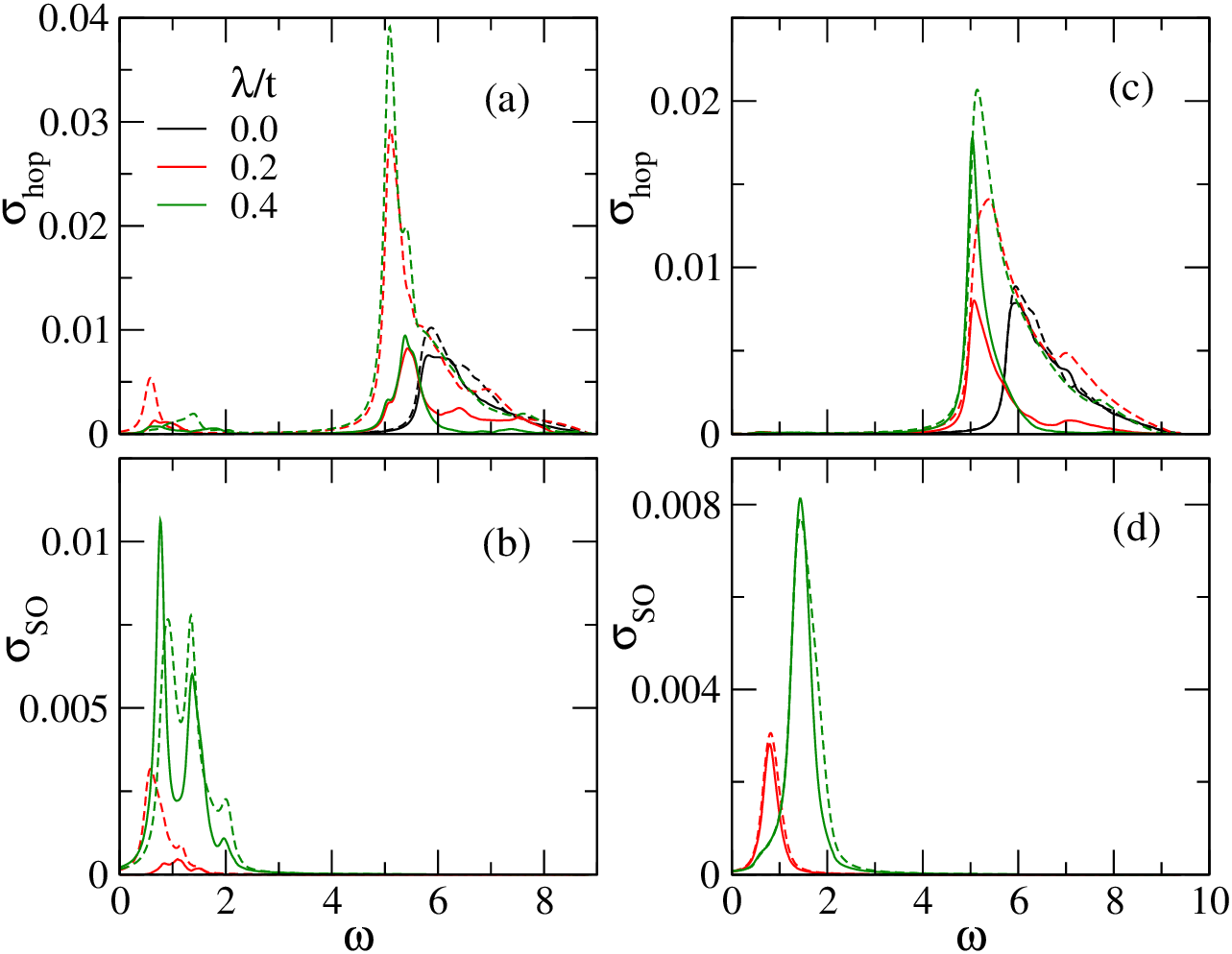}
\caption{(Color online) Optical conductivity as a function of the
frequency for Rashba strips coupled to AFM layers, $J_{sd}=5$ with
(a), (b) $W=4$, and (c), (d) $W=32$. The inter- and intra-band 
contributions to the part of $\sigma(\omega)$ due solely to
hopping currents are shown in (a), (c), while the ones to
$\sigma(\omega)$ due solely to the SO currents, in (b), (d).
Values of $\lambda/t$ are indicated on the plot. In all plots,
the interband (intraband) contribution is shown with dashed
(full) lines.}
\label{fig9}
\end{figure}

To understand the previous results for the optical conductivity,
let us now show the frequency dependence of its various contributions.
Fig.~\ref{fig9}(a,c) shows inter- and intra-band contributions to
the part of $\sigma(\omega)$ due solely to hopping currents,
$\sigma_{hop}(\omega)$, for $W=4$ and $W=32$ respectively, while
Fig.~\ref{fig9}(b,d), shows inter- and intra-band contributions to the
purely SO term,
$\sigma_{SO}(\omega)$, for $W=4$ and $W=32$ respectively. In the first
place, consistently with the behavior shown in Fig.~\ref{fig6}(c), the
intensity of the largest peaks in $\sigma_{hop}$ is larger than the
ones in $\sigma_{SO}$. Second, $\sigma_{hop}$ is mostly of interband
transitions, while $\sigma_{SO}$ has contributions from both inter-
and intraband transitions, and these contributions become approximately
equal particularly as $W$ and $\lambda/t$ increase, similarly to what
happens for isolated Rashba strips. It is quite 
apparent also that $\sigma_{hop}$, which is absent in the cases of
isolated Rashba strips and strips coupled to a FM layer shown in 
Fig.~\ref{fig5opt}, presents the strongest peak located at
$\omega \approx J_{sd}$, while $\sigma_{SO}$ has its strongest peaks
at low frequencies, and this behavior is again similar to the one
observed for $J_{sd}$.

\begin{figure}[t]
\includegraphics[width=0.9\columnwidth,angle=0]{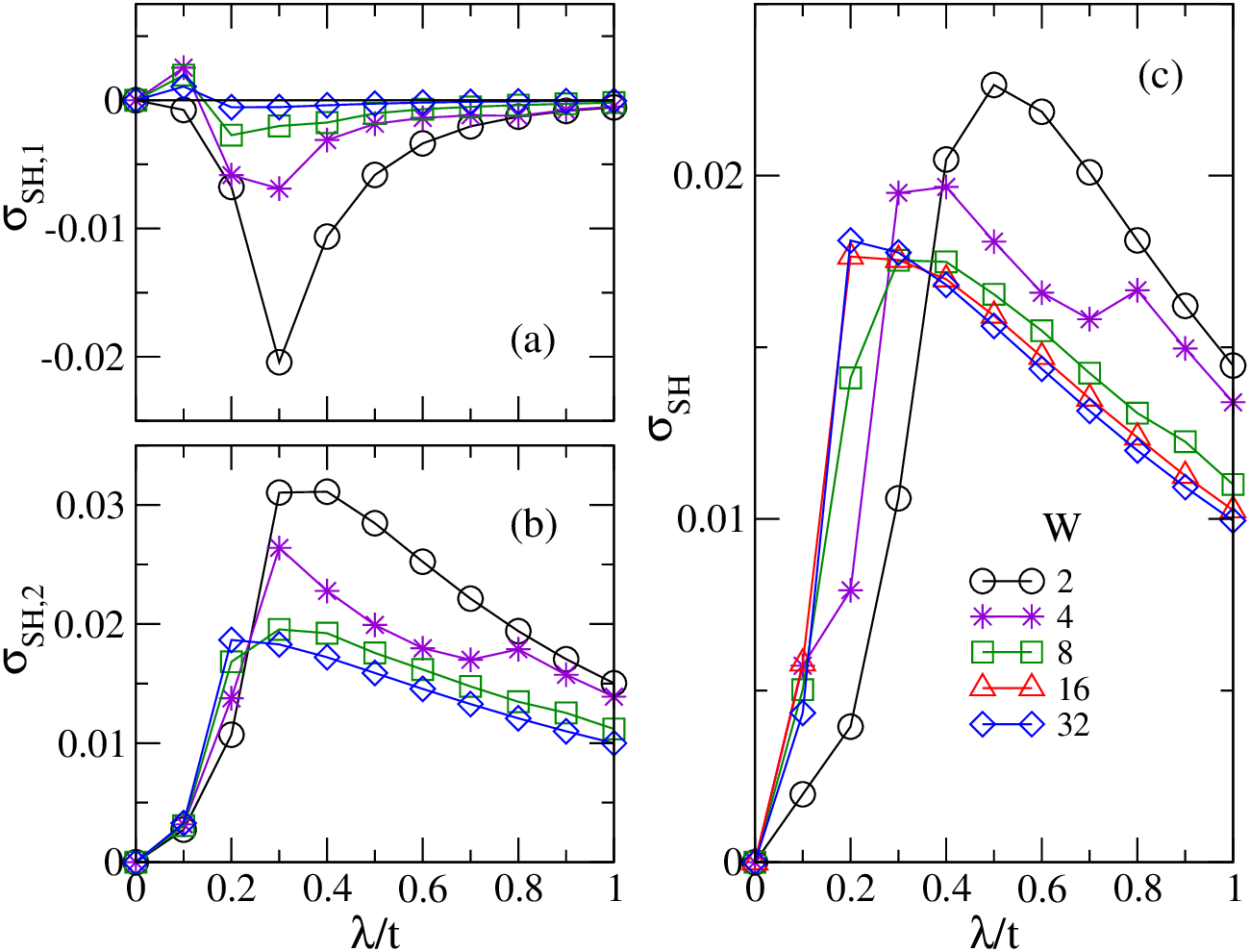}
\caption{(Color online) (a) $\sigma_{sH,1}$, and (b) $\sigma_{sH,2}$
contributions to the spin-Hall conductivity (see text), and (c)
total spin-Hall conductivity, as a function of $\lambda/t$,
for coupled AFM layer, $J_{sd}=5$, 
with symbols for various $W$ indicated on the plot.}
\label{fig10}
\end{figure}

Perhaps the most important and interesting results of the present
effort are the ones for the spin-Hall conductivity.
For the same reasons mentioned above, the spin-Hall conductivity will
also have contributions from both inter- and intraband processes.
However, what may be more interesting, is the fact that $\sigma_{sH}$
has contributions from both $<\hat{j}_{hop,x}~\hat{j}^s_{hop,y}>$
and $<\hat{j}_{SO,x}~\hat{j}^s_{hop,y}>$ terms. Contributions from
the former lead to the term $\sigma_{sH,1}$, while contributions from
the latter, to the term $\sigma_{sH,2}$. Results for $\sigma_{sH,1}$
and $\sigma_{sH,2}$ are shown in Figs.~\ref{fig10}(a,b), respectively.
The first conclusion is that both contributions have opposite signs,
except for $\lambda/t\le 1$, where both contributions are negative.
It is also apparent that $\sigma_{sH,1}$ is suppressed faster with
$W$. The total $\sigma_{sH}$, shown in Fig.~\ref{fig10}(c) is hence
dominated by the $<\hat{j}_{SO,x}~\hat{j}^s_{hop,y}>$ terms.
The discontinuous behavior above discussed is also noticeable 
in Figs.~\ref{fig10}(a,b), where a clear change of behavior can be 
seen in $0.2 \le \lambda/t \le 0.3$, that is coincidentally with
the peak in the RHCs shown in Fig.~\ref{fig2}(b).

The resulting behavior of $\sigma_{sH}$, shown in
Fig.~\ref{fig10}(c) is strikingly different to the one for FM coupled
layers. While for the FM layer, as shown in Fig.~\ref{fig5}(c),
$\sigma_{sH}$ increases with both $\lambda/t$ and $W$, for the AFM
case, $\sigma_{sH}$ reaches a maximum that
decreases with $W$, and this maximum is located at a value of
$\lambda/t$ that also decreases with $W$. The overall behavior 
of $\sigma_{sH}$ is also similar to the one already reported for
isolated Rashba strips (Fig.~\ref{fig4}(a)), including their
magnitude.

Last but not least, it is remarkable that the maximum value of
$\sigma_{sH}$ in the AFM background is more than two orders of magnitude
larger than the one for the FM case for the same $W$ and at the same
value of $\lambda/t$, and this difference is even larger for
narrower strip widths. The discussion of the interband or intraband
character of the transitions leading to $\sigma_{sH}$ are deferred
to the next Section.

\section{Other orientations of the coupled magnetic moments}
\label{afm_ins}

Let us assume that the magnetic moments of the coupled layer
can have an arbitrary but uniform orientation, forming an
angle $\theta$ with respect to the $z$-axis and an azimuthal angle
$\varphi$ with the $x$-axis.

Of course, in the absence of a Rashba SO coupling, the system is
independent of $\theta$, $\varphi$ but for any non-zero $\lambda$,
the physical properties will depend on the global orientation of the
magnetic moments. Actually, by minimizing the ground state energy of
Hamiltonian (\ref{ham1orb}), it turns out that the lowest energy
state with uniform orientation of the magnetic moments, for the FM
case, corresponds to $\theta=\pi/2$, $\varphi=0$, that is,
pointing along the $x$-axis, in the regime of large $J_{sd}$,
due to essentially a reduction of the $J_{sd}$ exchange energy.
The value of $J$ is irrelevant since for uniform orientation of
magnetic moments it would contribute to a constant independent of
$\lambda$.

Alternatively, the magnetic moment orientation may be fixed by the 
structure of the materials involved in a given device, so it is
interesting to examine the dependence of transport properties
with different orientations. Resuming the comment written after 
Eq.~(\ref{spcurz}), notice that now due to the $x$-component of
the magnetic moments, both $\hat{s}^z$ and the occupation number
operators, $n_\sigma$, do not commute with $H_{int}$. However,
the resulting contributions from these commutators are local,
field-like operators, no current-like ones. Hence, in the following,
the previous expressions for charge and spin currents, given by
Eq.~(\ref{current}) and Eq.~(\ref{spincury}), respectively, will
be used to compute the optical and spin-Hall conductivities.

\begin{figure}[t]
\includegraphics[width=0.9\columnwidth,angle=0]{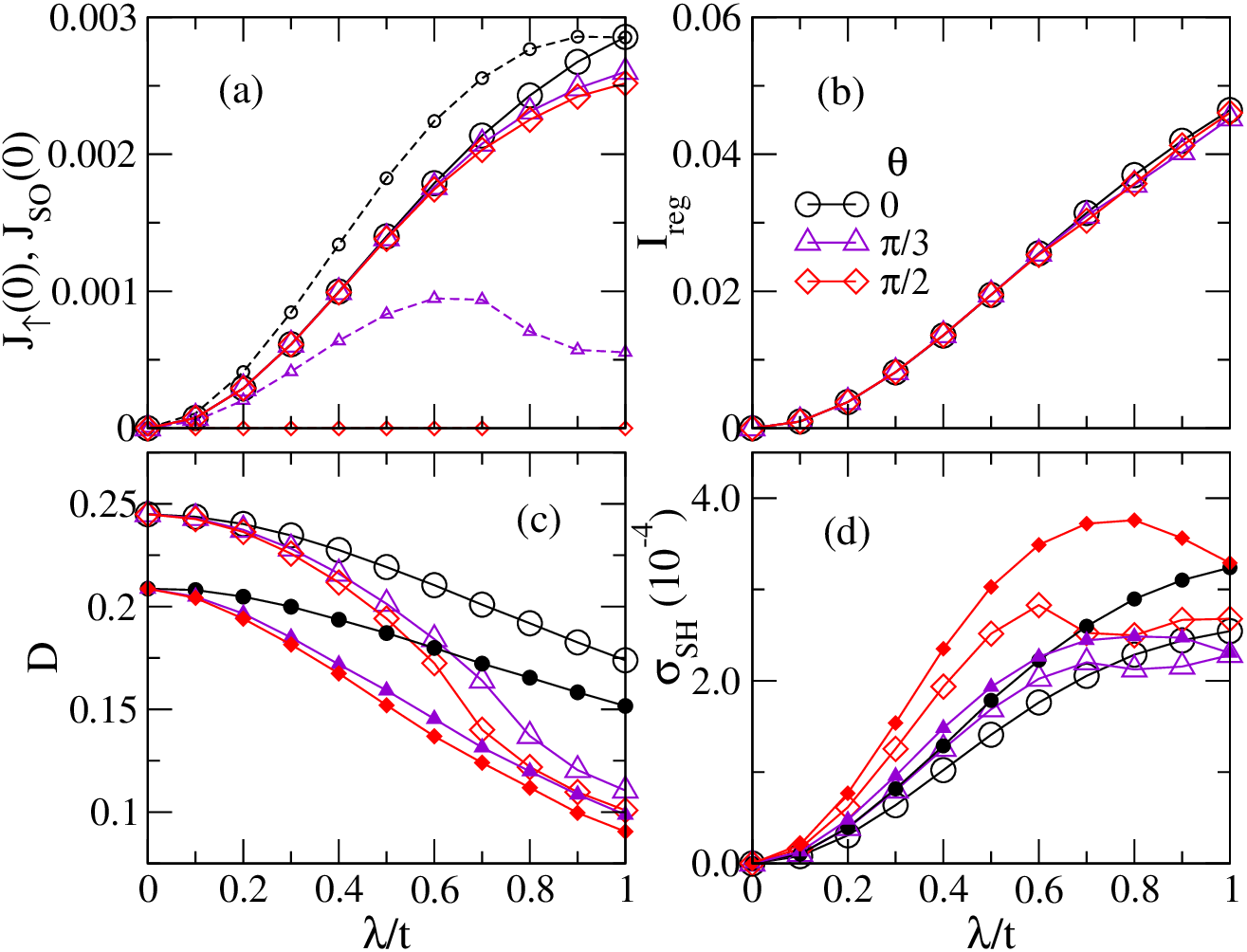}
\caption{(Color online) (a) Rashba helical current at the edge,
both of hopping (full lines) and SO (dashed lines) character,
(b) integral of the regular part of the optical conductivity, (c)
Drude weight, and (d) spin-Hall conductivity, as a function of
$\lambda/t$, for coupled FM layer, $J_{sd}=10$, $W=4$, and various
orientations of the magnetic moments, $\theta$ indicated on the
figure ($\varphi=0$). In (c) and (d) results for $W=32$ (full 
symbols) have been included.}
\label{fig11}
\end{figure}

In the following, two different global orientations of the magnetic
moments, $\theta=\pi/3$ and $\pi/2$, with $\varphi=0 $, will be
analyzed. Results for these two global orientations of the magnetic
moments for the FM layer, $J_{sd}=10$, are shown in Fig.~\ref{fig11},
together with previous results for $\theta=\varphi=0$ that are
included for comparison. Most of these results were obtained for
$W=4$ but are qualitatively independent of the width, as it can be
seen in Fig.~\ref{fig11}(c),(d), where the corresponding results
for $W=32$ were added. Fig.~\ref{fig11}(a) shows that the
RHCs of hopping origin at the strip's edge, $J_{\uparrow}(0)$,
only slightly decrease as $\theta$ increases from 0 to $\pi/2$,
but the RHCs of SO origin, which are nonzero for the FM layer as
discussed before, are strongly suppressed as $\theta$ is increased
from zero, and vanish for $\theta=\pi/2$, that is when the 
magnetization is parallel to the strip plane. This is consistent
with the previous comment after Fig.~\ref{fig5bis} about that the
RHC-SO reverse their direction when the magnetic moments are
inverted.

$I_{reg}$, shown in Fig.~\ref{fig11}(b), remains virtually unchanged
when $\theta$ changes from 0 to $\pi/2$. Its SO origin, discussed
in Section~\ref{coupledfm}, is also not modified by $\theta$.
There is a slight increase in the intraband contribution to $I_{reg}$,
but it still is much smaller than the interband one. Hence, mainly due
to a reduction in the kinetic energy, it can be seen in
Fig.~\ref{fig11}(c), that the Drude weight is homogeneously suppressed
by increasing $\theta$. It is also interesting to notice the 
enhancement of the spin-Hall conductivity for $\lambda/t \lesssim 0.7$
for $W=4$ ($\lambda/t\lesssim 0.7$ for $W=32$) as $\theta$ is increased
from 0 to $\pi/2$, as it can be observed in Fig.~\ref{fig11}(d). This
enhancement occurs even though the
$<\hat{j}_{SO,x}~\hat{j}^s_{hop,y}>$ origin of $\sigma_{sH}$, as well
as its purely interband character, are not modified by $\theta$.

\begin{figure}[t]
\includegraphics[width=0.9\columnwidth,angle=0]{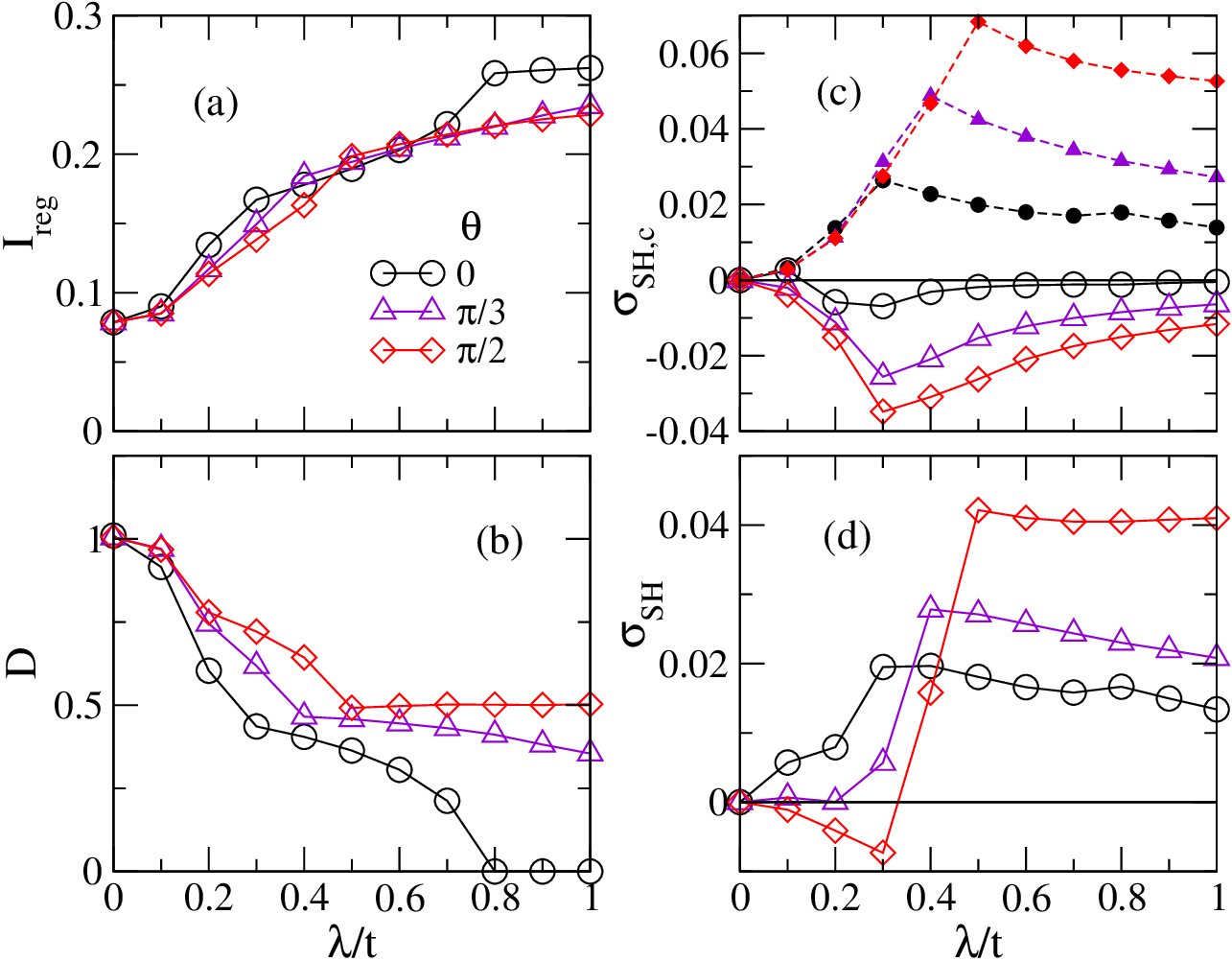}
\caption{(Color online) (a) Integral of the regular part of the
optical conductivity, (b) Drude weight, (c) $\sigma_{sH,1}$ (full
lines) and $\sigma_{sH,2}$ (dashed lines) contributions to the
spin-Hall conductivity (see text), and (d)
total spin-Hall conductivity, as a function of $\lambda/t$,
for coupled AFM layer, $J_{sd}=5$, $W=4$, and various orientations 
of the magnetic moments, $\theta$, indicated on the figure. The
azimuthal angle $\varphi=0$.}
\label{fig12}
\end{figure}

Let us now consider the also interesting case of the AFM coupled 
layer. In this case, the direction of the magnetic moments is
uniformly tilted with the constraint that magnetic moments on
nearest neighbor sites are kept antiparallel, that is, maintaining
the AFM order. In this case, as a difference with the FM layer,
if the antiparallel magnetic moments are freely rotated, the minimum
of the total energy of the Hamiltonian would correspond to 
$\theta=\varphi=0$, that is, they would point along the $z$-axis.

Results for these two global orientations of the magnetic moments,
for the $W=4$ strip, $J_{sd}=5$, are shown in Fig.~\ref{fig12},
together with previous results for $\theta=\varphi=0$
which are included for comparison. Fig.~\ref{fig12}(a) shows that
$I_{reg}$ does not depend significantly on
$\theta$. On the other hand, the Drude weight systematically increases
with $\theta$, particularly for $\lambda/t> 0.5$, as shown in
Fig.~\ref{fig12}(b), and this behavior can be attributed to the
closing of the gaps shown in Fig.~\ref{fig6}(a). However, even
for $\theta=\pi/2$, a change in the curvature can still be clearly
observed. The total spin-Hall conductivity is shown in
Fig.~\ref{fig12}(d). At small values of the RSOC, $\sigma_{sH}$ 
decreases with $\theta$, while for larger values of the RSOC, it
increases with $\theta$. It is remarkable that for small values of
$\lambda/t$, $\sigma_{sH}$ has an opposite signs for $\theta=0$ and
$\theta=\pi/2$. This behavior could be traced to the 
behavior of the two nonzero contributions to $\sigma_{sH}$, shown
in Fig.~\ref{fig12}(c). For small $\lambda/t$, the $\sigma_{sH,1}$
contribution, due to the $<\hat{j}_{hop,x}~\hat{j}^s_{hop,y}>$ terms,
grows rapidly with $\theta$, while $\sigma_{sH,2}$, due to the
$<\hat{j}_{SO,x}~\hat{j}^s_{hop,y}>$ terms, is virtually
independent of $\theta$. For large $\lambda/t$,
$\sigma_{sH,2}$ grows in absolute value more rapidly than
$\sigma_{sH,1}$, thus leading to the observed change in the 
behavior of the total $\sigma_{sH}$. 

\begin{figure}[t]
\includegraphics[width=0.9\columnwidth,angle=0]{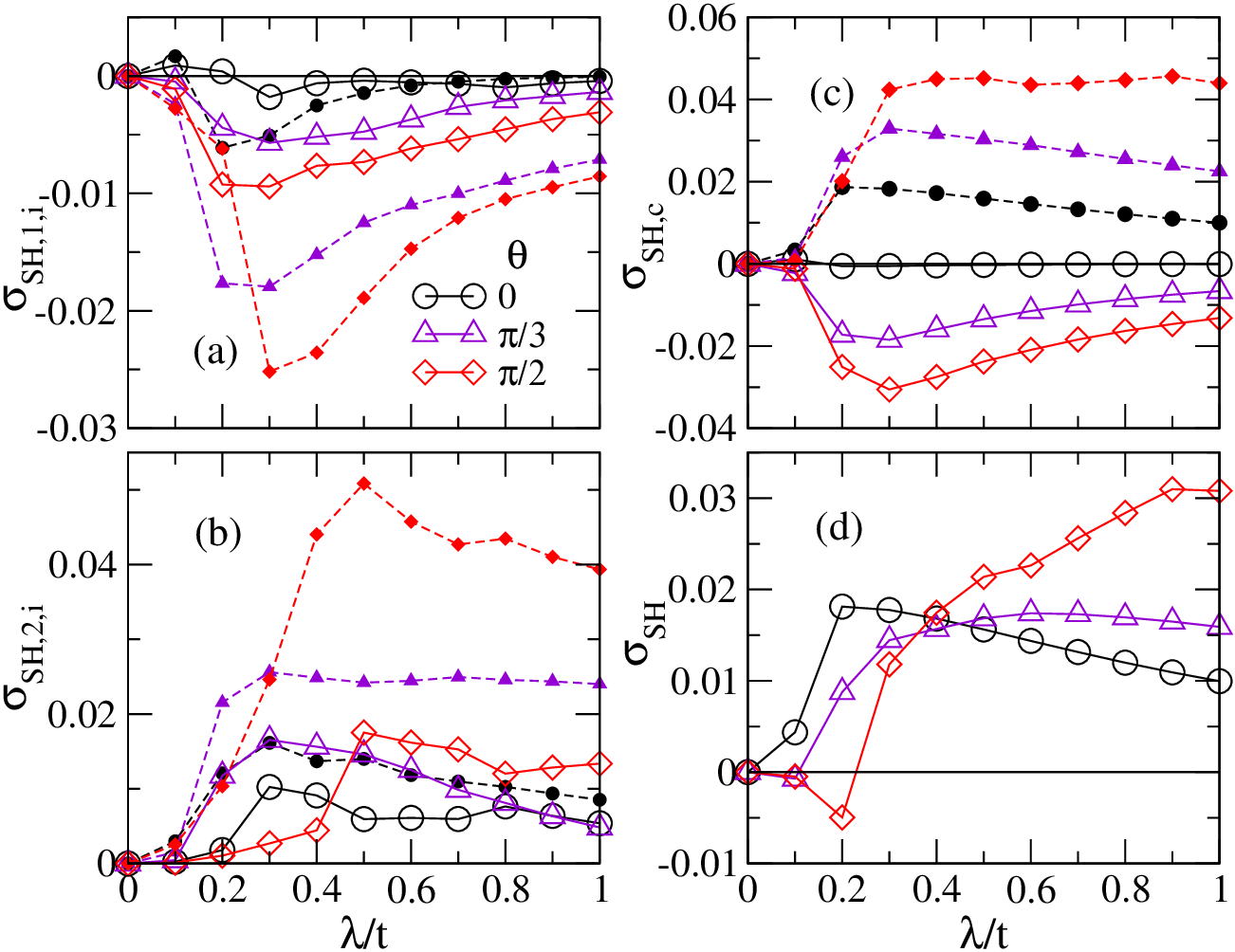}
\caption{(Color online) Intra- (full lines) and inter-band (dashed
lines) contributions to (a) $\sigma_{sH,1}$ and (b) $\sigma_{sH,2}$
(see text) as a function of $\lambda/t$, for $W=4$, and various
orientations of the magnetic moments, $\theta$, indicated on the
plot, $\varphi=0$.
(c) $\sigma_{sH,1}$ (full lines) and $\sigma_{sH,2}$ (dashed lines)
contributions to the spin-Hall conductivity, and (d) total spin-Hall
conductivity, as a function of $\lambda/t$, $W=32$.
Coupled AFM layer, $J_{sd}=5$.}
\label{fig13}
\end{figure}

The inter- or intra-band character of the transitions leading to 
$\sigma_{sH}$ is shown in Fig.~\ref{fig13}. Results depicted in
Fig.~\ref{fig13}(a), correspond to the inter- and intra-band
contributions to $\sigma_{sH,1}$ (shown with full lines in
Fig.~\ref{fig12}(c)), while the ones in Fig.~\ref{fig13}(b),
are for $\sigma_{sH,2}$ (shown with dashed lines in 
Fig.~\ref{fig12}(c)). In both cases, for virtually all values
of $\lambda/t$, and for the three values of $\theta$ considered,
the interband transitions are clearly dominant, but both inter-
and intra-band contributions increase with $\theta$. The previous
results, as well as those of Fig.~\ref{fig12}, correspond to
the $W=4$ strip. In Fig.~\ref{fig13}(c), the two contributions
$\sigma_{sH,1}$ and $\sigma_{sH,2}$, are shown for the $W=32$
strip. As in Fig.~\ref{fig12}(c), there is an increasing 
contribution from $<\hat{j}_{hop,x}~\hat{j}^s_{hop,y}>$
transitions, with an opposite sign to the 
$<\hat{j}_{SO,x}~\hat{j}^s_{hop,y}>$, which in absolute value are
in all cases dominant. The total $\sigma_{sH}$ for $W=32$ is
shown in Fig.~\ref{fig13}(d), and similar behaviors as those 
previously pointed out for $W=4$ are also present, particularly
the sign change for small $\lambda/t$ as $\theta$ increases from
0 to $\pi/2$.

If individual magnetic moments are allowed to rotate in order to
minimize the ground state energy of Hamiltonian (\ref{ham1orb}), it
is found that the AFM order is unstable towards an AFM spiral, for
$J\lessapprox 0.5$.
In this AFM spiral or double-spiral, the orientation of the magnetic
moments in one sublattice is given by $\theta= \kappa x$, and the
magnetic moments on the other sublattice have the opposite directions
thus keeping locally the AFM order.\cite{meza2}

For $J_{sd}=5$ or smaller, the AFM order is also unstable towards an
order that is staggered along the $y$-direction and spiral along the
$x$-direction, with $\theta= \kappa x$, $\kappa$ varying linearly
between 0 for $\lambda/t=0$, and $\pi/2$ for $\lambda/t \approx 0.8$.
This instability of the AFM order, driven by the conduction electrons,
disappears for $J\gtrapprox 0.5$.

The FM order for $J_{sd} > 5$ is also unstable towards an order that is
uniform along the transversal direction and spiral along the longitudinal 
direction again with a pitch $\kappa$ varying linearly between 0 and
$\pi/2$ as in the previously mentioned staggered spiral order, which is
stable for $J_{sd} < 5$. The electronically driven spiral instability
of the FM layer has been already reported,\cite{kim2013} as mentioned
above, but its dependence with $J_{sd}$, $\lambda/t$ and $W$, has not
been fully studied to the author's knowledge. A systematic study of
these various parameters on FM spiral orders will
be presented in a forthcoming study.\cite{meza2}

\section{Conclusions}

In this work, the effects of finite strip widths on transport properties
of Rashba conducting strips coupled to layers with various magnetic
orders have been studied numerically at zero temperature.

In the first place, it was found that for the uncoupled Rashba strip
(or coupled to a non-magnetic layer), the Drude weight slightly
decreases with the Rashba SO coupling, and has a weak dependence with
$W$. For a Rashba strip coupled to a FM layer, the Drude weight also
slightly decreases with $\lambda/t$ but it is strongly reduced by
increasing the strip width. In both cases, the Rashba SO coupling 
reduces the Drude weight through processes involving solely the SO
or spin-flipping currents. In addition, in the case of the FM
coupling, the processes contributing to the optical conductivity are
of interband origin, that is, connecting states with opposite signs
of $s^y$. In contrast, for the uncoupled Rashba strips, these 
processes are completely of interband nature for the narrowest strip
with $W=2$, but the contribution from intraband transitions starts to
grow with increasing $W$ until it becomes approximately equal to the 
one of interband transitions, in the whole range of $\lambda/t$
examined.

For the AFM coupled layer, due to the opening of gaps in the 
single-particle spectrum, the Drude weight has a more complex
behavior. For all strip widths and Rashba SO couplings considered,
its value is in general much smaller than for the other two cases
studied, and it is much strongly suppressed by $\lambda/t$. This
suppression is due both to processes involving spin-conserving and
spin-flipping currents. For the narrowest strips, $W=2,~4$, $D$ even
vanishes at some finite value of $\lambda/t$. Certainly, this 
different behavior with respect to the other two cases, which are
metallic, indicates the proximity to an insulating behavior in the
AFM case.

Of course, the most relevant results for spintronic applications are
those concerning the spin-Hall conductivity. In this sense, the main
result is that $\sigma_{sH}$ in the coupled AFM system is nearly two
orders of magnitude larger than the one in the coupled FM system for
virtually all values of $W$ and $\lambda/t$. It would be tempting to
relate this result with the recent finding of N\'eel-spin orbit
torques driving domain walls at velocities two orders of magnitude
greater than the ones in ferromagnets,\cite{gomonay,zelezny16} but
this feature belongs to off-equilibrium regimes.

It is also important to emphasize that, for strips coupled to an
AFM layer, a larger value of $\sigma_{sH}$ is achieved for narrow
(wide) strips at large (small) values of the Rashba SO coupling.
The crossover between both regimes occurs at a value of $\lambda/t$
that approximately coincides with the value at which a peak
appears on the edge helical currents. Actually, the behavior of
$\sigma_{sH}$ with $W$ and $\lambda/t$ is quite similar to the one
obtained for uncoupled Rashba strips. However, in the uncoupled case,
$\sigma_{sH}$ is entirely due to processes involving charge 
SO currents along $x$ and spin hopping currents along $y$, while 
in the AFM case, there are also contributions from processes
involving spin hopping currents along $y$, although these 
contributions have opposite sign to the former ones, except for
$\lambda/t < 0.1$.

The orientation of the magnetic moments for the AFM coupled
layer\cite{zelezny16} has also been examined. It is interesting to
note that for $W=4$, $\sigma_{sH}$ is largest at small $\lambda/t$ when
the magnetic moments are oriented along the $z$-direction, while for
magnetic moments oriented along the $x$-axis, which is the case studied
in Ref.~\onlinecite{sinova14}, $\sigma_{sH}$ is largest at large
$\lambda/t$.

Finally, in systems where the conducting-magnetic layers exchange
$J_{sd}$ is due to a Hund coupling between conducting and localized
orbitals of a transition metal oxide, which may be the case of
devices involving SrTiO$_3$ interfaces, the ordering of the magnetic
moments is determined by the dynamics of the competing degrees of
freedom. In these systems, the AFM order become unstable at small
and intermediate values of $\lambda/t$ with
respect to a double spiral order, where both the Drude weight and
$\sigma_{sH}$ essentially vanish for all $W$.

\begin{acknowledgments}
Useful discussions with C. Gazza, I. Hamad, and G. Meza, 
are gratefully acknowledged.
The author is partially supported by the Consejo Nacional de
Investigaciones Cient\'ificas y T\'ecnicas (CONICET) of Argentina
through grant PIP No. 11220120100389CO.
\end{acknowledgments}


\begin{thebibliography}{}
\bibitem{note} In addition, $[H_{int,y,l},\hat{s}^z_l]=0$, if the
       direction of the magnetic moments is along the $z$-direction.
\end{thebibliography}
\end{document}